\documentclass{lncs/svproc}
\usepackage{xspace}
\usepackage[dvipsnames]{xcolor}
\usepackage{afterpage}
\usepackage{figlatex}
\usepackage{listings,mathtools}
\usepackage{mathrsfs}
\usepackage{multirow}
\usepackage{enumitem}
\usepackage[ruled,vlined,linesnumbered,noend]{algorithm2e}
\usepackage{framed}
\usepackage{wrapfig}
\usepackage{amsmath}
\usepackage{amssymb}

\usepackage[colorlinks]{hyperref}
\hypersetup{
	colorlinks = true,
	citecolor = {Orange},
	linkcolor = {Magenta},
	urlcolor  = {blue}
}

\usepackage{smartdiagram}
\usepackage{tikz}
\usetikzlibrary{arrows,positioning}

\setlist[itemize]{noitemsep, topsep=0pt}
\newcommand{\tab}{\quad\quad}

\newcommand{\definedas}{\triangleq\xspace}
\newcommand{\ie}{{\em i.e.}\xspace}
\newcommand{\st}{\ \mbox{s.t.}\ }

\renewcommand{\^}{\xspace\wedge\xspace}
\renewcommand{\v}{\xspace\vee\xspace}

\newcommand{\union}{\xspace\cup\xspace}

\newcommand{\nin}{\not\in\xspace}

\newcommand{\bigo}{\mathcal{O}\xspace}

\newcommand{\fp}{\texttt{FP}\xspace}

\newcommand{\hl}[1]{\hypertarget{#1}{\textcolor{darkgray}{\texttt{#1}}}\xspace} 
\newcommand{\hlref}[1]{\hyperlink{#1}{\textcolor{Sepia}{\small \texttt{(#1)}}}}

\newcommand{\cnf}{\texttt{CNF}\xspace}
\newcommand{\sat}{\texttt{SAT}\xspace}

\newcommand{\ourtechniquename}{FenSying}
\newcommand{\ourfasttechniquename}{fFenSying}
\newcommand{\ourtechnique}{\textcolor{MidnightBlue}{\texttt{\ourtechniquename}}\xspace}

\newcommand{\fasttechnique}{\textcolor{MidnightBlue}{\texttt{\ourfasttechniquename}}\xspace}
\newcommand{\cc}{\textit{\small C11}\xspace}
\newcommand{\cds}{CDSChecker\xspace}

\newcommand{\z}{\texttt{Z3}\xspace}
\newcommand{\stfence}{\textcolor{black}{Strong-\ourtechniquename}\xspace}
\newcommand{\wkfence}{\textcolor{black}{Weak-\ourtechniquename}\xspace}
\newcommand{\btg}{\texttt{\small BTG}\xspace}

\newcommand{\inv}[1]{{#1}^{\mathtt{inv}}\xspace}
\newcommand{\imm}[1]{{#1}^{\mathtt{imm}}\xspace}
\newcommand{\fx}[1]{{#1}^{\mathtt{fx}}\xspace}
\newcommand{\formula}[1]{\mathcal{Q}(#1)\xspace}

\newcommand{\moset}{\mathcal{M}\xspace}

\newcommand{\objects}{\mathcal{O}\xspace}

\newcommand{\wt}[1]{{\mathbb{W}#1}}
\newcommand{\rd}[1]{{\mathbb{R}#1}}
\newcommand{\fn}[1]{{\mathbb{F}#1}}
\newcommand{\events}{\mathcal{E}\xspace}
\newcommand{\writes}{\events^\wt{}\xspace}
\newcommand{\reads}{\events^\rd{}\xspace}
\newcommand{\fences}{\events^\fn{}\xspace}

\newcommand{\ordevents}[1]{\events^{(#1)}\xspace}
\newcommand{\ordwrites}[1]{\events^\wt{(#1)}\xspace}
\newcommand{\ordreads}[1]{\events^\rd{(#1)}\xspace}
\newcommand{\ordfences}[1]{\events^\fn{(#1)}\xspace}

\newcommand{\molt}{{\sqsubset}\xspace}
\newcommand{\mole}{{\sqsubseteq}\xspace}
\newcommand{\mogt}{{\sqsupset}\xspace}
\newcommand{\moge}{{\sqsupseteq}\xspace}

\newcommand{\reln}[4]{#3 {\rightarrow^{#1}_{#2}} #4\xspace} 

\newcommand{\seqb}[3]{\reln{\textbf{\textcolor{CarnationPink}{sb}}}{#1}{#2}{#3}\xspace}
\newcommand{\rf}[3]{\reln{\textbf{\textcolor{PineGreen}{rf}}}{#1}{#2}{#3}\xspace} 
 
\newcommand{\dob}[3]{\reln{\textbf{\textcolor{Mulberry}{dob}}}{#1}{#2}{#3}\xspace}

\newcommand{\sw}[3]{\reln{\textbf{\textcolor{Magenta}{sw}}}{#1}{#2}{#3}\xspace}

\newcommand{\ithb}[3]{\reln{\textbf{\textcolor{NavyBlue}{ithb}}}{#1}{#2}{#3}\xspace}

\newcommand{\hb}[3]{\reln{\textbf{\textcolor{Cerulean}{hb}}}{#1}{#2}{#3}\xspace}

\newcommand{\mo}[3]{\reln{\textbf{\textcolor{RedOrange}{mo}}}{#1}{#2}{#3}\xspace}

\renewcommand{\to}[3]{\reln{\textbf{\textcolor{Brown}{to}}}{#1}{#2}{#3}\xspace}

\newcommand{\so}[3]{\reln{\textbf{\textcolor{Mahogany}{so}}}{#1}{#2}{#3}\xspace} 
\newcommand{\fr}[3]{\reln{\textbf{\textcolor{RoyalPurple}{fr}}}{#1}{#2}{#3}} 

\newcommand{\rfinv}[3]{#2 \rightarrow^{ \textcolor{PineGreen}{ \textbf{rf}}^{-1} }_{#1} #3\xspace}

\newcommand{\setSB}{\seqb{\tau}{}{}\xspace}
\newcommand{\setRF}{\rf{\tau}{}{}\xspace}
\newcommand{\setSW}{\sw{\tau}{}{}\xspace}
\newcommand{\setDOB}{\dob{\tau}{}{}\xspace}
\newcommand{\setITHB}{\ithb{\tau}{}{}\xspace}
\newcommand{\setHB}{\hb{\tau}{}{}\xspace}
\newcommand{\setMO}{\mo{\tau}{}{}\xspace}
\newcommand{\setTO}{\to{\tau}{}{}\xspace}
\newcommand{\setSO}{\so{\tau}{}{}\xspace}
\newcommand{\setFR}{\fr{\tau}{}{}\xspace}

\newcommand{\isetSB}{\seqb{\imm{\tau}}{}{}\xspace}
\newcommand{\isetRF}{\rf{\imm{\tau}}{}{}\xspace}
\newcommand{\isetSW}{\sw{\imm{\tau}}{}{}\xspace}
\newcommand{\isetDOB}{\dob{\imm{\tau}}{}{}\xspace}
\newcommand{\isetITHB}{\ithb{\imm{\tau}}{}{}\xspace}
\newcommand{\isetHB}{\hb{\imm{\tau}}{}{}\xspace}
\newcommand{\isetMO}{\mo{\imm{\tau}}{}{}\xspace}
\newcommand{\isetTO}{\to{\imm{\tau}}{}{}\xspace}
\newcommand{\isetSO}{\so{\imm{\tau}}{}{}\xspace}
\newcommand{\isetFR}{\fr{\imm{\tau}}{}{}\xspace}

\newcommand{\lsb}{\textbf{\textcolor{CarnationPink}{sb}}\xspace}
\newcommand{\lrf}{\textbf{\textcolor{PineGreen}{rf}}\xspace}
\newcommand{\ldob}{\textbf{\textcolor{Mulberry}{dob}}\xspace}
\newcommand{\lsw}{\textbf{\textcolor{Magenta}{sw}}\xspace}

\newcommand{\lhb}{\textbf{\textcolor{Cerulean}{hb}}\xspace}
\newcommand{\lmo}{\textbf{\textcolor{RedOrange}{mo}}\xspace}
\newcommand{\lto}{\textbf{\textcolor{Brown}{to}}\xspace}
\newcommand{\lso}{\textbf{\textcolor{Mahogany}{so}}\xspace}
\newcommand{\lfr}{\textbf{\textcolor{RoyalPurple}{fr}}\xspace}

\newcommand{\na}{\texttt{na}\xspace}
\newcommand{\rlx}{\texttt{rlx}\xspace}
\newcommand{\rel}{\texttt{rel}\xspace}
\newcommand{\acq}{\texttt{acq}\xspace}
\newcommand{\acqrel}{\texttt{ar}\xspace}
\renewcommand{\sc}{\texttt{sc}\xspace}
\newcommand{\onsc}[1]{{#1{|}_\sc} \xspace}

\newcommand{\fto}{\textcolor{BrickRed}{FTo}\xspace}
\newcommand{\bto}{\textcolor{Emerald}{BTo}\xspace}

\newcommand{\SOtr}[3]{\reln{\textbf{\textcolor{Mahogany}{so}}^+}{#1}{#2}{#3}}
\newcommand{\isetSOtr}{\SOtr{\imm{\tau}}{}{}}
\usepackage{url}

\begin{document}
\mainmatter              
\title{Fence Synthesis under the C11 Memory Model}
\titlerunning{Fence Synthesis under the C11 Memory Model}  
%
\author{Sanjana Singh\inst{1} \and Divyanjali Sharma\inst{1}
Ishita Jaju\inst{2} \and Subodh Sharma\inst{1}}
\authorrunning{S. Singh, D. Sharma, I. Jaju, S. Sharma} 
%
%
\institute{Indian Institute of Technology Delhi, India,\\
\email{$\{$sanjana.singh,divyanjali,svs$\}$@cse.iitd.ac.in},\\ 
\and
Uppsala University, Sweden}

\maketitle              

\begin{abstract}
The C/C++11 (\cc) standard offers a spectrum of ordering guarantees 
on memory access operations. The combinations of such orderings 
pose a challenge in  developing  {\em correct} and 
{\em efficient} weak memory programs.
A common solution to preclude those program outcomes that violate the
correctness specification is using \cc synchronization-fences,
which establish ordering on program events.
The challenge is in choosing a combination of fences that 
(i) restores the correctness of the input program, with
(ii) as little impact on efficiency as possible
(\ie, the smallest set of weakest fences). 
This problem is the {\em optimal fence synthesis}
problem and is NP-hard for straight-line programs.
In this work, we propose the first fence synthesis technique for \cc
programs called \ourtechnique and show its optimality.  We
additionally propose a near-optimal efficient alternative called
\fasttechnique.  We prove the optimality of \ourtechnique and the soundness of
\fasttechnique and present an implementation of both techniques.
Finally, we contrast the performance of the two techniques and
empirically demonstrate \fasttechnique's effectiveness.

\keywords{\cc, fence-synthesis, optimal}
\end{abstract}

\section{Introduction} \label{sec:intro}
%
Developing weak memory programs requires careful placement of
fences and memory barriers to preserve ordering between program 
instructions and exclude undesirable program outcomes. 
However, computing the correct combination of the 
type and location of fences is challenging.
Too few or incorrectly placed fences may not preserve the necessary 
ordering, while too many fences can negatively impact the performance.
Striking a balance between preserving the correctness and obtaining
performance is highly non-trivial even for expert programmers.
%

This paper presents an automated fence synthesis solution for
weak memory programs developed using the C/C++11 standard (\cc).
\cc provides a spectrum of ordering guarantees called {\em memory 
orders}. In a program, a memory access operation is associated 
with a memory order which specifies how other memory accesses 
are ordered with respect to the operation.
The memory orders range from {\em relaxed} (\rlx) (that imposes no 
ordering restriction) to {\em sequentially-consistent} (\sc) (that
may restore sequential consistency).
Understanding all the subtle complexities of \cc orderings and
predicting the program outcomes can quickly become exacting.
Consider the program \hlref{RWRW} ($\mathsection$\ref{sec:overview}),
where the orders are shown as subscripts.
When all the memory accesses are ordered
\rlx, there exists a program outcome that violates the
correctness specification (specified as an {\em assert}
statement). However, when all accesses are ordered \sc, the
program is provably correct.

%
\raggedbottom
In addition, the \cc memory model supports \cc {\em fences} that serve
as tools for imposing ordering restrictions. 
Notably, \cc associates fences with memory orders, thus, 
supporting various degrees of ordering guarantees through fences.
%
%
%
%
%

This work proposes an {\em optimal} fence synthesis technique for \cc
called \ourtechnique. It involves finding solutions to two problems:
(i) computing an optimal (minimal) set of locations to synthesize fences and
(ii) computing an optimal (weakest) memory order to be associated with the 
fences (formally defined in $\mathsection$\ref{sec:preliminaries}).
\ourtechnique takes as input {\em all} program runs that 
violate user-specified assertions and attempts optimal \cc fence 
synthesis to stop the violating outcomes.
\ourtechnique reports when \cc fences alone cannot fix a violation.
In general, computing a minimal number of fences with multiple types of 
fences is shown to be NP-hard for straight-line 
programs~\cite{taheri2019polynomial}.
We note,
rather unsurprisingly, that this hardness manifests in the proposed
optimal fence synthesis solution even for the simplest \cc
programs. Our experiments ($\mathsection$\ref{sec:results}) show an
exponential increase in the analysis time with the increase in the
program size.

Further, to address scalability, this paper proposes a {\em
near-optimal} fence synthesis technique called \fasttechnique
(\textcolor{MidnightBlue}{\tt f}ast \ourtechnique) that fixes 
{\em one} violating outcome at a time optimally.
Note that fixing one outcome optimally may not 
guarantee optimality across all violating outcomes.
In the process, this technique may add a small number of extra fences
than what an optimal solution would compute.
Our experiments reveal that \fasttechnique performs exponentially 
better than \ourtechnique in terms of the analysis time while
adding no extra fences in over 99.5\% of the experiments.

Both \ourtechnique and \fasttechnique, compute the solution from a
 set of combinations of fences that can stop the violating outcomes,
 also called {\em candidate solutions}.  The candidate solutions
 are encoded in a {\em head-cycle-free} \cnf \sat 
 query~\cite{angiulli2014tractability}. Computing an
 optimal solution from candidates then becomes finding a solution to a
 {\em min-model finding problem}.

Many prior works have focused on automating fence synthesis
(discussed in $\mathsection$\ref{sec:related}).
%
%
However, the techniques presented in this paper are distinct from prior 
works in the following two ways: 
(i) prior techniques do not support C11 memory orders, and 
(ii) the proposed techniques in this paper synthesize fences that are 
portable and not architecture-specific.

\noindent{\bf Contributions.}
To summarize, this work makes the following contributions:
\begin{itemize}
\item The paper presents \ourtechnique
and \fasttechnique ($\mathsection$\ref{sec:methodology}).
To the best of our knowledge, these are the
first fence synthesis techniques for \cc.  
\item The paper shows (using Theorems~\ref{thm:fastfensying sound}
and~\ref{thm:fensying sound}) that the techniques are sound, \ie, if
the input program can be fixed by \cc fences, then the techniques will
indeed find a solution.  The paper also shows (using
Theorem~\ref{thm:fensying optimal}) that \ourtechnique produces an optimal
result in the number and type of fences.
\item Finally, the paper presents an implementation of the said techniques
and presents an empirical validation 
using a set of 1389 litmus tests. 
Further, the paper empirically shows the effectiveness
of \fasttechnique on a set of challenging benchmarks from prior works.
\fasttechnique performs on an average 67x faster than
\ourtechnique.
\end{itemize}
\section{Overview of \ourtechnique and \fasttechnique} \label{sec:overview}
%
%
%
%
\begin{figure}[!t]
	\centering
	\begin{minipage}{0.4\textwidth}
		\centering
		{\setlength{\tabcolsep}{5pt}
			\begin{tabular}{|l||l|}
				\hline
				\multicolumn{2}{|c|}{Initially: $x=0, y=0$} \\
				
				$ a :=_\sc y $   & $ b :=_\sc x $ \\
				$ x :=_\rlx 1 $ & $ y :=_\rlx 1 $ \\
				
				\multicolumn{2}{|c|}{assert($\neg(a{=}1 \^ b{=}1)$)} \\
				\hline
				
				\multicolumn{2}{c}{\hl{RWRW}}
		\end{tabular}}
		\begin{tabular}{|c|}
			\hline
			\resizebox{\textwidth}{!}{\tikzset{every picture/.style={line width=0.75pt}} 
\begin{tikzpicture}[x=1em,y=1em,yscale=-1,xscale=-1]
	\tikzstyle{every node}=[font=\normalfont]
	\node (ix) [inner sep=2pt,color=Brown] {$\mathbb{I}(x,0)$};
	\node (iy) [right=22pt of ix, inner sep=2pt,color=Brown] {$\mathbb{I}(y,0)$};
	\node (ry) [below left=5pt and -24pt of ix, inner sep=2pt] {$ R^\sc(y,1)$};
	\node (rx) [right=30pt of ry, inner sep=2pt] {$R^\sc(x,1)$};
	\node (wx) [below=40pt of ry, inner sep=2pt] {$W^\rlx(x,1)$};
	\node (wy) [below=40pt of rx, inner sep=2pt] {$W^\rlx(y,1)$};
	
	\node (f11) [above=15pt of ry, inner sep=2pt, ,color=White] {$\mathbb{F}_{11}$};
	\node (f12) [below=15pt of ry, inner sep=2pt, ,color=White] {$\mathbb{F}_{12}$};
	
	\node (f21) [above=15pt of rx, inner sep=2pt, ,color=White] {$\mathbb{F}_{21}$};
	\node (f22) [below=15pt of rx, inner sep=2pt, ,color=White] {$\mathbb{F}_{22}$};
	\draw [->,>=stealth,color=RedOrange] ($ (ix.south west)+(-0.1,-5pt) $) to[out=15,in=-60] node[pos=0.3,left=-1pt,font=\scriptsize,color=black] { $\lmo$ } ($ (wx.north west)+(0pt,0) $);
	\draw [->,>=stealth,color=RedOrange] ($ (iy.south east)+(-0.1,-3pt) $) to[out=165,in=-120] node[pos=0.3,right=-1pt,font=\scriptsize,color=black] { $\lmo$ } ($ (wy.north east)+(0pt,0) $);
	\draw [->,>=stealth,color=PineGreen] (wx) -- node[pos=0.8,above=-1pt,font=\scriptsize,color=black] { $\lrf$ } (rx);
	\draw [->,>=stealth,color=CarnationPink] (rx) -- node[midway,right=-2pt,font=\scriptsize,color=black] { $\lsb$ } (wy);
	\draw [->,>=stealth,color=PineGreen] (wy) -- node[pos=0.8,above=-1pt,font=\scriptsize,color=black] { $\lrf$ } (ry);
	\draw [->,>=stealth,color=CarnationPink] (ry) -- node[midway,left=-2pt,font=\scriptsize,color=black] { $\lsb$ } (wx);
	
	\draw [opacity=0.3] (-2.5,2) -- (-2.5,8);
	\draw [opacity=0.3] (-2.7,2) -- (-2.7,8);
	
\end{tikzpicture}} \\
			\hline
			\multicolumn{1}{c}{\hl{RWRW-bt}} \\
			
			\hline
			\resizebox{\textwidth}{!}{\tikzset{every picture/.style={line width=0.75pt}} 
\begin{tikzpicture}[x=1em,y=1em,yscale=-1,xscale=-1]
	\tikzstyle{every node}=[font=\normalfont]
	\node (ix) [inner sep=2pt,color=Brown] {$\mathbb{I}(x,0)$};
	\node (iy) [right=22pt of ix, inner sep=2pt,color=Brown] {$\mathbb{I}(y,0)$};
	\node (ry) [below left=30pt and -22pt of ix, inner sep=2pt] {$ R^\sc(y,1)$};
	\node (rx) [right=30pt of ry, inner sep=2pt] {$R^\sc(x,1)$};
	\node (wx) [below=40pt of ry, inner sep=2pt] {$W^\rlx(x,1)$};
	\node (wy) [below=40pt of rx, inner sep=2pt] {$W^\rlx(y,1)$};
	
	\node (f11) [above=15pt of ry, inner sep=2pt, ,color=Blue] {$\mathbb{F}_{11}$};
	\node (f12) [below=15pt of ry, inner sep=2pt, ,color=Blue] {$\mathbb{F}_{12}$};
	\node (f13) [below=15pt of wx, inner sep=2pt, ,color=Blue] {$\mathbb{F}_{13}$};
	
	\node (f21) [above=15pt of rx, inner sep=2pt, ,color=Blue] {$\mathbb{F}_{21}$};
	\node (f22) [below=15pt of rx, inner sep=2pt, ,color=Blue] {$\mathbb{F}_{22}$};
	\node (f23) [below=15pt of wy, inner sep=2pt, ,color=Blue] {$\mathbb{F}_{23}$};

	\draw [->,>=stealth,color=CarnationPink] (f11) -- node[midway,left=-2pt,font=\scriptsize,color=black] { $\lsb$ } (ry);
	\draw [->,>=stealth,color=CarnationPink] (ry) -- node[midway,left=-2pt,font=\scriptsize,color=black] { $\lsb$ } (f12);
	\draw [->,>=stealth,color=CarnationPink] (f12) -- node[midway,left=-2pt,font=\scriptsize,color=black] { $\lsb$ } (wx);
	\draw [->,>=stealth,color=CarnationPink] (wx) -- node[midway,left=-2pt,font=\scriptsize,color=black] { $\lsb$ } (f13);
	
	\draw [->,>=stealth,color=CarnationPink] (f21) -- node[midway,right=-2pt,font=\scriptsize,color=black] { $\lsb$ } (rx);
	\draw [->,>=stealth,color=CarnationPink] (rx) -- node[midway,right=-2pt,font=\scriptsize,color=black] { $\lsb$ } (f22);
	\draw [->,>=stealth,color=CarnationPink] (f22) -- node[midway,right=-2pt,font=\scriptsize,color=black] { $\lsb$ } (wy);
	\draw [->,>=stealth,color=CarnationPink] (wy) -- node[midway,right=-2pt,font=\scriptsize,color=black] { $\lsb$ } (f23);

	\draw [->,>=stealth,color=RedOrange] ($ (ix.south west)+(-0.1,-5pt) $) to[out=15,in=-60] node[pos=0.3,left=-1pt,font=\scriptsize,color=black] { $\lmo$ } ($ (wx.north west)+(0pt,0) $);
	\draw [->,>=stealth,color=RedOrange] ($ (iy.south east)+(-0.1,-3pt) $) to[out=165,in=-120] node[pos=0.3,right=-1pt,font=\scriptsize,color=black] { $\lmo$ } ($ (wy.north east)+(0pt,0) $);
	\draw [->,>=stealth,color=PineGreen] (wx) -- node[pos=0.8,above=-1pt,font=\scriptsize,color=black] { $\lrf$ } (rx);
	\draw [->,>=stealth,color=PineGreen] (wy) -- node[pos=0.8,above=-1pt,font=\scriptsize,color=black] { $\lrf$ } (ry);

	\draw [opacity=0.3] (-2.5,2) -- (-2.5,13);
	\draw [opacity=0.3] (-2.7,2) -- (-2.7,13);
	
\end{tikzpicture}} \\
			\hline
			\multicolumn{1}{c}{\hl{RWRW-imm}} \\
		\end{tabular}
	\end{minipage}
\hspace{2em}
	\begin{minipage}{0.4\textwidth}
		\begin{tabular}{|c|}					
			\hline
			\resizebox{\textwidth}{!}{\tikzset{every picture/.style={line width=0.75pt}} 
\begin{tikzpicture}[x=1em,y=1em,yscale=-1,xscale=-1]
	\tikzstyle{every node}=[font=\normalfont]
	\node (ix) [inner sep=2pt,color=Brown] {$\mathbb{I}(x,0)$};
	\node (iy) [right=22pt of ix, inner sep=2pt,color=Brown] {$\mathbb{I}(y,0)$};
	\node (ry) [below left=5pt and -24pt of ix, inner sep=2pt] {$ R^\sc(y,1)$};
	\node (rx) [right=30pt of ry, inner sep=2pt] {$R^\sc(x,1) $};
	\node (wx) [below=40pt of ry, inner sep=2pt] {$W^\rlx(x,1)$};
	\node (wy) [below=40pt of rx, inner sep=2pt] {$W^\rlx(y,1)$};
	
	\node (f11) [above=15pt of ry, inner sep=2pt, ,color=White] {$\mathbb{F}_{11}$};
	\node (f12) [below=15pt of ry, inner sep=2pt, ,color=Blue] {$\mathbb{F}_{12}^\sc$};
	
	\node (f21) [above=15pt of rx, inner sep=2pt, ,color=White] {$\mathbb{F}_{21}$};
	\node (f22) [below=15pt of rx, inner sep=2pt, ,color=Blue] {$\mathbb{F}_{22}^\sc$};

	\draw [->,>=stealth,color=CarnationPink,opacity=0.0] (ry) -- node[midway,left=-2pt,font=\scriptsize,color=black] { $\lsb$ } (f12);
	\draw [->,>=stealth,dashed,color=CarnationPink,opacity=0.7] (f12) -- node[midway,left=-2pt,font=\scriptsize,color=black] { $\lsb$ } (wx);
	
	\draw [->,>=stealth,color=CarnationPink,opacity=0.0] (rx) -- node[midway,right=-2pt,font=\scriptsize,color=black] { $\lsb$ } (f22);
	\draw [->,>=stealth,dashed,color=CarnationPink,opacity=0.7] (f22) -- node[midway,right=-2pt,font=\scriptsize,color=black] { $\lsb$ } (wy);
	
	\draw [->,>=stealth,dashed,color=RedOrange,opacity=0.7] ($ (ix.south west)+(-0.1,-5pt) $) to[out=15,in=-60] node[pos=0.3,left=-1pt,font=\scriptsize,color=black] { $\lmo$ } ($ (wx.north west)+(0pt,0) $);
	\draw [->,>=stealth,dashed,color=RedOrange,opacity=0.7] ($ (iy.south east)+(-0.1,-3pt) $) to[out=165,in=-120] node[pos=0.3,right=-1pt,font=\scriptsize,color=black] { $\lmo$ } ($ (wy.north east)+(0pt,0) $);
	\draw [->,>=stealth,dashed,color=PineGreen,opacity=0.0] (wx) -- node[pos=0.8,above=-1pt,font=\scriptsize,color=black] { $\lrf$ } (rx);
	\draw [->,>=stealth,dashed,color=PineGreen,opacity=0.0] (wy) -- node[pos=0.8,above=-1pt,font=\scriptsize,color=black] { $\lrf$ } (ry);
	
	\draw [->,>=stealth,color=Brown] (f12)  -- node[pos=0.2,above=-1pt,font=\scriptsize,color=black] { $\lto$ } (rx);
	\draw [->,>=stealth,color=Brown] (rx)  -- node[midway,right=-2pt,font=\scriptsize,color=black] { $\lto$ } (f22);
	\draw [->,>=stealth,color=Brown] (f22)  -- node[pos=0.2,above=-1pt,font=\scriptsize,color=black] { $\lto$ } (ry);
	\draw [->,>=stealth,color=Brown] (ry)  -- node[midway,left=-2pt,font=\scriptsize,color=black] { $\lto$ } (f12);
	
	\draw [opacity=0.3] (-2.5,2) -- (-2.5,8);
	\draw [opacity=0.3] (-2.7,2) -- (-2.7,8);
	
\end{tikzpicture}} \\
			\hline
			\multicolumn{1}{c}{\hl{RWRW-inv-to}} \\	
			
			\hline
			\resizebox{\textwidth}{!}{\tikzset{every picture/.style={line width=0.75pt}} 
\begin{tikzpicture}[x=1em,y=1em,yscale=-1,xscale=-1]
	\tikzstyle{every node}=[font=\normalfont]
	\node (ix) [inner sep=2pt,color=Brown] {$\mathbb{I}(x,0)$};
	\node (iy) [right=22pt of ix, inner sep=2pt,color=Brown] {$\mathbb{I}(y,0)$};
	\node (ry) [below left=5pt and -24pt of ix, inner sep=2pt] {$ R^\sc(y,1)$};
	\node (rx) [right=30pt of ry, inner sep=2pt] {$R^\sc(x,1) $};
	\node (wx) [below=40pt of ry, inner sep=2pt] {$W^\rlx(x,1)$};
	\node (wy) [below=40pt of rx, inner sep=2pt] {$W^\rlx(y,1)$};
	
	\node (f11) [above=15pt of ry, inner sep=2pt, ,color=White] {$\mathbb{F}_{11}$};
	\node (f12) [below=15pt of ry, inner sep=2pt, ,color=Blue] {$\mathbb{F}_{12}^\rel$};
	
	\node (f21) [above=15pt of rx, inner sep=2pt, ,color=White] {$\mathbb{F}_{21}$};
	\node (f22) [below=15pt of rx, inner sep=2pt, ,color=Blue] {$\mathbb{F}_{22}^\acq$};

	\draw [->,>=stealth,color=CarnationPink] (ry) -- node[midway,left=-2pt,font=\scriptsize,color=black] { $\lsb$ } (f12);
	\draw [->,>=stealth,color=CarnationPink,opacity=1.0] (f12) -- node[midway,left=-2pt,font=\scriptsize,color=black] { $\lsb$ } (wx);
	
	\draw [->,>=stealth,color=CarnationPink,opacity=1.0] (rx) -- node[midway,right=-2pt,font=\scriptsize,color=black] { $\lsb$ } (f22);
	\draw [->,>=stealth,color=CarnationPink] (f22) -- node[midway,right=-2pt,font=\scriptsize,color=black] { $\lsb$ } (wy);
	
	\draw [->,>=stealth,dashed,color=RedOrange,opacity=0.7] ($ (ix.south west)+(-0.1,-5pt) $) to[out=15,in=-60] node[pos=0.3,left=-1pt,font=\scriptsize,color=black] { $\lmo$ } ($ (wx.north west)+(0pt,0) $);
	\draw [->,>=stealth,dashed,color=RedOrange,opacity=0.7] ($ (iy.south east)+(-0.1,-3pt) $) to[out=165,in=-120] node[pos=0.3,right=-1pt,font=\scriptsize,color=black] { $\lmo$ } ($ (wy.north east)+(0pt,0) $);
	\draw [->,>=stealth,dashed,color=PineGreen,opacity=0.0] (wx) -- node[pos=0.8,above=-1pt,font=\scriptsize,color=black] { $\lrf$ } (rx);
	\draw [->,>=stealth,dashed,color=PineGreen,opacity=0.0] (wy) -- node[pos=0.8,above=-1pt,font=\scriptsize,color=black] { $\lrf$ } (ry);
	
	\draw [->,>=stealth,color=Cerulean] (f12) -- node[pos=0.7,above=-2pt,font=\scriptsize,color=black] { $\lsw$ } node[pos=0.7,below=-2pt,font=\scriptsize,color=black] { $\lhb$ } (f22);

	
	\draw [opacity=0.3] (-2.5,2) -- (-2.5,8);
	\draw [opacity=0.3] (-2.7,2) -- (-2.7,8);
	
\end{tikzpicture}} \\
			\hline
			\multicolumn{1}{c}{\hl{RWRW-inv-sync}} \\
			
			\hline
			\resizebox{\textwidth}{!}{\tikzset{every picture/.style={line width=0.75pt}} 
\begin{tikzpicture}[x=1em,y=1em,yscale=-1,xscale=-1]
	\tikzstyle{every node}=[font=\normalfont]
	\node (ix) [inner sep=2pt,color=Brown] {$\mathbb{I}(x,0)$};
	\node (iy) [right=22pt of ix, inner sep=2pt,color=Brown] {$\mathbb{I}(y,0)$};
	\node (ry) [below left=5pt and -24pt of ix, inner sep=2pt] {$ R^\sc(y,1)$};
	\node (rx) [right=30pt of ry, inner sep=2pt] {$R^\sc(x,1) $};
	\node (wx) [below=40pt of ry, inner sep=2pt] {$W^\rlx(x,1)$};
	\node (wy) [below=40pt of rx, inner sep=2pt] {$W^\rlx(y,1)$};
	
	\node (f11) [above=15pt of ry, inner sep=2pt, ,color=White] {$\mathbb{F}_{11}$};
	\node (f12) [below=15pt of ry, inner sep=2pt, ,color=Blue] {$\mathbb{F}_{12}^\rel$};
	
	\node (f21) [above=15pt of rx, inner sep=2pt, ,color=White] {$\mathbb{F}_{21}$};
	\node (f22) [below=15pt of rx, inner sep=2pt, ,color=White] {$\mathbb{F}_{22}$};

	\draw [->,>=stealth,color=CarnationPink,opacity=1.0] (ry) -- node[midway,left=-2pt,font=\scriptsize,color=black] { $\lsb$ } (f12);
	\draw [->,>=stealth,color=CarnationPink,opacity=1.0] (f12) -- node[midway,left=-2pt,font=\scriptsize,color=black] { $\lsb$ } (wx);
	
	\draw [->,>=stealth,color=CarnationPink,opacity=1.0] (rx) -- node[midway,right=-2pt,font=\scriptsize,color=black] { $\lsb$ } (wy);
	
	\draw [->,>=stealth,dashed,color=RedOrange,opacity=0.7] ($ (ix.south west)+(-0.1,-5pt) $) to[out=15,in=-60] node[pos=0.3,left=-1pt,font=\scriptsize,color=black] { $\lmo$ } ($ (wx.north west)+(0pt,0) $);
	\draw [->,>=stealth,dashed,color=RedOrange,opacity=0.7] ($ (iy.south east)+(-0.1,-3pt) $) to[out=165,in=-120] node[pos=0.3,right=-1pt,font=\scriptsize,color=black] { $\lmo$ } ($ (wy.north east)+(0pt,0) $);
	\draw [->,>=stealth,color=PineGreen,opacity=0.0] (wx) -- node[pos=0.8,below=-1pt,font=\scriptsize,color=black] { $\lrf$ } (rx);
	\draw [->,>=stealth,color=PineGreen,opacity=0.0] (wy) -- node[pos=0.8,above=-1pt,font=\scriptsize,color=black] { $\lrf$ } (ry);
	
	\draw [->,>=stealth,color=Cerulean] (f12)  -- node[pos=0.7,above=-1pt,font=\scriptsize,color=black] { $\lsw$ } node[pos=0.7,below=-1pt,font=\scriptsize,color=black] { $\lhb$ } (rx);
	
	\draw [opacity=0.3] (-2.5,2) -- (-2.5,8);
	\draw [opacity=0.3] (-2.7,2) -- (-2.7,8);
	
\end{tikzpicture}} \\
			\hline
			\multicolumn{1}{c}{\hl{RWRW-inv-sync-opt}} \\
		\end{tabular}
	\end{minipage}
\end{figure}
Given a program $P$,
a {\em trace} $\tau$ of $P$ (formally defined in 
$\mathsection$\ref{sec:preliminaries}); 
is considered {\em buggy} if it violates an assertion of $P$.
\ourtechnique takes {\em all} buggy traces of $P$ 
as input.  The difference in \fasttechnique is that the input is a 
{\em single} buggy trace of $P$.
%

%


%

Consider the input program \hlref{RWRW}, where $x$ and $y$ are shared
objects with initial values 0, and $a$ and $b$ are local objects.  Let
$W^m(o,v)$ and $R^m(o,v)$ represent the write and read of object $o$
and value $v$ with the memory order $m$. Let $\mathbb{I}(o,v)$
represent the initialization event for object $o$ with value $v$.  The
parallel bars ($\parallel$) represent the parallel composition of
events from separate threads.
Figure~\hlref{RWRW-bt} represents a buggy trace $\tau$ of
\hlref{RWRW} under \cc semantics.  For convenience, the relations $-$
{\em sequenced-before} ($\setSB$), {\em reads-from} ($\setRF$), 
{\em modification-order} ($\setMO$) (formally defined in 
$\mathsection$\ref{sec:preliminaries},$\mathsection$\ref{sec:c11}) $-$ 
among the events of $\tau$ are shown. 
The assert condition of \hlref{RWRW} is violated as 
the read events are not ordered before the write events of the same object, 
allowing reads from {\em later} writes.
%
%

Consider the following three sets of fences that can
invalidate the trace \hlref{RWRW-bt}:
$c_1$ = $\{\mathbb{F}^\sc_{12}, \mathbb{F}^\sc_{22}\}$,
$c_2$ = $\{\mathbb{F}^\rel_{12}, \mathbb{F}^\acq_{22}\}$ and 
$c_3$ = $\{\mathbb{F}^\rel_{12}\}$
(the superscripts indicate the memory orders and the
subscripts  represent the synthesis locations for fences).
The  solutions are depicted in Figures \hlref{RWRW-inv-to}, 
\hlref{RWRW-inv-sync} and \hlref{RWRW-inv-sync-opt} for $c_1$,
$c_2$ and $c_3$, respectively.
The candidate solution $c_1$  prevents a total order 
on the \sc ordered events \cite{batty2011mathematizing,C11},
thus, invalidating \hlref{RWRW-inv-to} under \cc semantics.
With candidate solution $c_2$, a  happens-before ($\setHB$) ordering 
is formed  (refer $\mathsection$\ref{sec:c11}) between $R^\sc(y,1)$ 
and $W^\sc(y,1)$. 
This forbids a read from an ordered-later write, thus, invalidating
\hlref{RWRW-inv-sync}.
Candidate solution $c_3$ establishes a similar $\setHB$ ordering by 
exploiting the strong memory order of $R^\sc(x,1)$ and invalidates 
\hlref{RWRW-inv-sync-opt}.

The candidate solution $c_2$ is preferred over $c_1$ as it contains
weaker fences. On the other hand, candidate $c_3$ represents an
optimal solution as it uses the smallest number of weakest fences.
We formally define the optimality of fence synthesis in
$\mathsection$\ref{sec:preliminaries}.
While \ourtechnique will compute the solution $c_3$, 
\fasttechnique may compute one from the many candidate solutions.

Both \ourtechnique and \fasttechnique start by transforming each
buggy trace $\tau$ to an {\em intermediate} version, $\imm{\tau}$,
by inserting {\em untyped} \cc fences (called {\em candidate fences}) 
above and below the trace events. \hlref{RWRW-imm} shows an intermediate 
version corresponding to \hlref{RWRW-bt}.
%
The addition of fences (assuming they are of the strongest variety)
leads to the creation of new $\setHB$ ordering edges. This may
result in cycles in the dependency graph under the
\cc semantics (refer $\mathsection$\ref{sec:c11}). 
The set
of fences in a cycle constitutes a {\em candidate solution}.
For example, an ordering from $\mathbb{F}_{12}$ to $\mathbb{F}_{22}$ 
in \hlref{RWRW-imm} induces a cyclic 
relation $W^\rlx(y,1)$ $\setRF$ $R^\sc(y,1)$ $\setHB$ $W^\rlx(y,1)$
violating the $\setRF;\setHB$ irreflexivity 
(refer $\mathsection$\ref{sec:c11}).
%
%

The candidate solutions are collected in a \sat query 
($\Phi$). Assuming $c_1$, $c_2$ and $c_3$ 
are the only candidate solutions for \hlref{RWRW-bt}, then $\Phi$ = 
($\mathbb{F}_{12}$ $\^$ $\mathbb{F}_{22}$) $\v$
($\mathbb{F}_{12}$ $\^$ $\mathbb{F}_{22}$) $\v$ ($\mathbb{F}_{12}$),
where for a fence $\mathbb{F}^m_i$, $\mathbb{F}_i$ represents the 
same fence with unassigned memory order.
 \fasttechnique
uses a \sat solver to compute the {\em min-model} of $\Phi$, {\tt min}$\Phi$ = 
$\{\mathbb{F}_{12}\}$.
Further,
\fasttechnique applies the \cc ordering rules on fences 
 to determine the weakest memory order for the fences in {\tt min}$\Phi$.
For instance,  $\mathbb{F}_{12}$ in {\tt min}$\Phi$ is computed to have
the order \rel (refer $\mathsection$\ref{sec:methodology}).
\fasttechnique then inserts $\mathbb{F}_{12}$ with memory order \rel
in \hlref{RWRW} at the location depicted in
\hlref{RWRW-inv-sync-opt}. This process repeats for the next buggy
trace.

In contrast, since \ourtechnique works with all buggy traces at once,
it requires the conjunction of the \sat queries $\Phi_i$ corresponding
to each buggy trace $\tau_i$. The min-model of the conjunction is
computed, which provides optimality.
%


%
\section{Preliminaries} \label{sec:preliminaries}
Consider a multi-threaded \cc program ($P$). Each thread of $P$ 
performs a sequence of {\em events} that are
runtime instances of memory access operations (reads, writes, and rmws) on
shared objects and \cc fences.
Note that an event is uniquely identified in a trace; however,
multiple events may be associated with the same program location.
The events may be atomic or non-atomic.
Given the set $\objects$ of shared  objects accessed by $P$
and the set $\moset$ of \cc memory orders, an event is formally
defined as:
{\definition [Event] \label{def:event}}
An event $e$ is a tuple 
$\langle thr(e)$, $idx(e)$ $act(e)$, $obj(e)$, $ord(e)$, $loc(e) 
\rangle$ where,
\begin{itemize}[label=act(e),align=left,leftmargin=4em]
\item [$thr(e)$] $\in P$, represents the thread of $e$;
\item [$idx(e)$] represents the identifier of $e$ unique to $thr(e)$;
\item [$act(e)$] $\in \{$read, write, rmw ({\em read-modify-write}), 
	fence$\}$, is the event action;
\item [$obj(e)$] $\subseteq \objects$, is the set of memory objects 
	accessed by $e$;
\item [$ord(e)$] $\subseteq \moset$, is the \cc memory order associated 
	with $e$; and
\item [$loc(e)$] is the program location corresponding to $e$.
\end{itemize}
%
%

\noindent
{\bf \cc memory orders.}
The atomic events and fence operations are associated with memory
orders that define the ordering restriction on atomic and non-atomic 
events around them.
Let $\moset$ = $\{$\na, \rlx, \rel, \acq, \acqrel, $\sc\}$, 
represent the orders relaxed (\rlx), release (\rel), 
acquire$/$consume (\acq), acquire-release (\acqrel) and 
sequentially consistent (\sc) for atomic events. 
A non-atomic event is recognized by the \na memory order.
Let $\molt \subseteq \moset \times \moset$ represent the relation
{\em weaker} such that $m_1 \molt m_2$ represents that the  $m_1$ is 
weaker than $m_2$. As a consequence, annotating an event with
$m_2$ may order two events that remain unordered with $m_1$. 
The orders in $\moset$ are related as
\na $\molt$ \rlx $\molt$ $\{\rel,\acq\}$ $\molt$ \acqrel $\molt$ \sc.
We also define the relation $\mole$ to represent {\em weaker or equally
weak}. 
Similarly, we define $\mogt$ to represent {\em stronger} 
and $\moge$ to represent {\em stronger or equally strong}.

%
We use $\writes \subseteq \events$ to denote the set 
of events that perform write to shared memory objects \ie, events 
$e$ \st $act(e)$ = write or rmw. Similarly, we use $\reads \subseteq 
\events$ to denote events $e$ that read from a shared memory object
\ie, $act(e)$ = read or rmw, and $\fences$ to denote the fence events
$e$ \st $act(e)$ = fence.
We also use $\ordevents{m}\in \events$ (and accordingly 
$\ordwrites{m}$, $\ordreads{m}$ and $\ordfences{m}$) to 
represent the events with the memory order $m \in \moset$;
as an example $\ordfences{\sc}$ represents the set of fences
with the memory order \sc. 
{\definition [Trace] \label{def:trace}}
A {\em trace}, $\tau$, of $P$  is a tuple 
$\langle \events_\tau, \setHB, \setMO, \setRF \rangle$, 
where
\begin{itemize}[label=sethb,align=left,leftmargin=*]
	\item [$\events_\tau$] $\subseteq \events$
		represents the set of events in the trace 
		$\tau$;
	\item [$\setHB$] ({\em Happens-before}) 
		$\subseteq \events_\tau \times \events_\tau$
		is a partial order  which captures the event 
		interactions and inter-thread synchronizations, 
		discussed  in $\mathsection$\ref{sec:c11};
	\item [$\setMO$] ({\em Modification-order})
		$\subseteq \writes_\tau \times \writes_\tau$
		is a total order on the writes of an object;
	\item [$\setRF$] ({\em Reads-from}) 
		$\subseteq \writes_\tau \times \reads_\tau$
		is a relation from a write event to a read event signifying 
		that the read event takes its value from the write event in 
		$\tau$.
\end{itemize}

Note that, we use $\writes_\tau$, $\reads_\tau$ and $\fences_\tau$
(and also $\ordwrites{m}_\tau$, $\ordreads{m}_\tau$ and 
$\ordfences{m}_\tau$ where $m \in \moset$)
for the respective sets of events for a trace $\tau$.

\noindent
{\bf Relational Operators.}
$R^{-1}$ represents the inverse and $R^+$ represents the 
transitive closure of a relation $R$.
Further, $R_1;R_2$ represents the composition of relations 
$R_1$ and $R_2$.
Let $\onsc{R}$ represent a subset of a relation $R$ on \sc 
ordered events; \ie $(e_1,e_2) \in \onsc{R}$ $\iff$ $(e_1,e_2) 
\in R$ $\^$ $e_1,e_2 \in \ordevents{\sc}$. Note that we 
also use the infix notation $e_1Re_2$ for $(e_1,e_2) \in R$.
Lastly, a relation $R$ has a cycle (or is cyclic) if 
$\exists e_1, e_2 \in \events$ \st $e_1Re_2$ $\^$ $e_2Re_1$.


\noindent
{\bf A note on optimality.}
The notion of optimality may vary with context.
Consider two candidate solutions $\{\mathbb{F}_i^\sc\}$
 and $\{\mathbb{F}^\rel_j, \mathbb{F}^\acq_k\}$ where the superscripts 
 represent the memory orders. The two solutions are incomparable
 under \cc, and their performance efficiency is subject to the
 input program and the underlying architecture.
\ourtechnique chooses a candidate solution $c$ as an optimal solution 
if:
(i)  $c$ has the smallest number of candidate fences, and 	
(ii) each fence of $c$ has the weakest memory order
 compared to other candidate solutions that satisfy (i).

Let $sz(c)$ represent the size of the candidate solution $c$ and given 
the set of all candidate solutions $\{c_1, ..., c_n\}$ to fix $P$,
let $\underline{sz}(P)$ = {\tt min}($sz(c_1)$, ..., $sz(c_n)$).
Further, we assign weights $wt(c)$ to each candidate solution $c$,
computed as the summation of the weights of its fences where
a fence ordered \rel or \acq is assigned the weight 1, a fence 
ordered \acqrel is assigned 2, and a fence ordered \sc is assigned 3. 
Optimality for \ourtechnique is formally defined as:
\begin{definition}{\bf Optimality of fence synthesis.}
	\normalfont
	Consider a set of candidate solutions $c_1, ..., c_n$.
	A solution $c_i$ (for $i \in [1,n]$) is considered optimal if:

	(i) $sz(c_i)$ = $\underline{sz}(P)$ $\^$
	(ii) $\forall j \in [1,n]$ \st $sz(c_j)$ = $\underline{sz}(P)$,
	$wt$($c_i$) $\le$ $wt$($c_j$).
	\label{def:optimality}
\end{definition}

\section{Background: C11 Memory Model} \label{sec:c11}
The \cc memory model defines a trace using a set of event 
relations, described in Definition~\ref{def:trace}. 
%
The most significant relation that defines a \cc trace $\tau$ 
is the irreflexive and acyclic happens-before relation,
$\setHB$ $\subseteq$ $\events_\tau \times \events_\tau$.
The $\setHB$ relation is composed of the following relations
\cite{C11}.
\begin{itemize}[label=setHB,align=left,leftmargin=*]
	\item [$\setSB$] ({\em Sequenced-before}): total occurrence order 
		on the events of a thread;
		\ie \newline
		$\forall e_1, e_2 \in \events_\tau$ \st $thr(e_1)$ = $thr(e_2)$
		and $e_1$ occurs before $e_2$ in their thread $\implies$
		$\seqb{\tau}{e_1}{e_2}$.
		
	\item [$\setSW$] ({\em Synchronizes-with}) Inter-thread 
		synchronization between a  write $e_w$ 
		(ordered $\moge$ \rel) and a  read $e_r$ 
		(ordered $\moge$ \acq) when $\rf{\tau}{e_w}{e_r}$;
		\ie \newline
		$\forall e_w \in \writes_\tau$, $e_r \in \reads_\tau$ \st
		$ord(e_w) \moge \rel$ and $ord(e_r) \moge \acq$,
		$\rf{\tau}{e_w}{e_r}$ $\implies$ $\sw{\tau}{e_w}{e_r}$.
		
	\item [$\setDOB$] ({\em Dependency-ordered-before}): Inter-thread
		synchronization between a  write $e_w$ 
		(ordered $\moge$ \rel) and a  read $e_r$ 
		(ordered $\moge$ \acq) when $\rf{\tau}{e_w'}{e_r}$ for
		$e_w' \in$ {\em release-sequence}\footnotemark of $e_w$
		in $\tau$ \cite{batty2011mathematizing,C11};
		\ie \newline
		$\forall e_w, e_w' \in \writes_\tau$, $e_r \in \reads_\tau$ \st
		$ord(e_w) \moge \rel$ and $ord(e_r) \moge \acq$,
		$e_w' \in$ {\em release-sequence} of $e_w$ and
		$\rf{\tau}{e_w'}{e_r}$ $\implies$ $\dob{\tau}{e_w}{e_r}$.
\footnotetext{{\em release-sequence} of $e_w$
	in $\tau$: maximal contiguous sub-sequence 
	of $\setMO$ that starts at $e_w$ and contains: (i) write events of $thr(e_w)$,
	(ii) rmw events of other threads \cite{batty2011mathematizing,C11}.}

	\item [$\setITHB$] ({\em Inter-thread-hb}): 
		Inter-thread relation computed by extending  $\setSW$ and 
		$\setDOB$ with $\setSB$;
		\ie \newline
		$\forall e_1, e_2, e_3 \in \events_\tau$,
		\begin{enumerate}[leftmargin=2em]
			\item $\sw{\tau}{e_1}{e_2}$, {\em or}
			\item $\dob{\tau}{e_1}{e_2}$, {\em or}
			\item $\sw{\tau}{e_1}{e_3}$ $\^$ $\seqb{\tau}{e_3}{e_2}$, {\em or}
			\item $\sw{\tau}{e_1}{e_3}$ $\^$ $\ithb{\tau}{e_3}{e_2}$, {\em or}
			\item $\ithb{\tau}{e_1}{e_3}$ $\^$ $\ithb{\tau}{e_3}{e_2}$
		\end{enumerate} 
		$\implies$ $\ithb{\tau}{e_1}{e_2}$.
	
	\item [$\setHB$] ({\em Happens-before}): 
		Inter-thread relation defined as 
		$\setSB$ $\union$ $\setITHB$.
\end{itemize}

The $\setHB$ relation 
along with the $\setMO$ and $\setRF$ relations
(Definition~\ref{def:trace}) is used in specifying 
the set of six \hl{coherence conditions}~\cite{C11,LahavVafeiadis-PLDI17}:

%
%

\begin{itemize}[align=left,leftmargin=3em]
	\item[] $\setHB$ is irreflexive. \hfill (\hl{co-h})
	\item[] $\setRF;\setHB$ is irreflexive. \hfill (\hl{co-rh})
	\item[] $\setMO;\setHB$ is irreflexive. \hfill (\hl{co-mh})
	\item[] $\setMO;\setRF;\setHB$ is irreflexive. \hfill (\hl{co-mrh})
	\item[] $\setMO;\setHB;\setRF^{-1}$ is irreflexive. \hfill (\hl{co-mhi})
	\item[] $\setMO;\setRF;\setHB;\setRF^{-1}$ is irreflexive. \hfill (\hl{co-mrhi})
\end{itemize}

Additionally, {\em all} \sc ordered events in a trace $\tau$ must be related
by a total order ($\setTO$) that concurs with the coherence conditions.
We use an irreflexive relation called {\em from-reads} 
($\setFR$ $\definedas$ $\setRF^{-1}$;$\setMO$) 
for ordering reads with {\em later} writes. 
%
%
%
Consequently, $\setTO$ must satisfy the following condition 
\cite{C11,vafeiadis2015common}: 

$order(P,R)$ $\definedas$ 
($\nexists a\ R(a,a)$) $\^$
($R^+ \subseteq R$) $\^$
($R \subseteq P \times P$);  and,

$total(P,R)$ $\ \definedas$
$\forall a,b \in P$ $\implies$
$a = b$ $\v$ $R(a,b)$ $\v$ $R(b,a)$.
\newline
All \sc ordered events must form a total order $\setTO$ \st
the following conditions are satisfied:

\begin{enumerate}
	\item $order(\ordevents{\sc},\setTO)$ $\^$ 
	$total(\ordevents{\sc},\setTO)$ $\^$ $\onsc{\setHB}$ 
	$\union$ $\onsc{\setMO}$ $\subseteq$ $\setTO$
	\hfill (\hl{coto})
	\item $\forall \rf{\tau}{e_w}{e_r}$ \st $e_r \in 
	\ordevents{\sc}_\tau$
	\begin{itemize}
		\item either, $e_w \in \ordevents{\sc}_\tau$ $\^$ 
		{\tt imm-scr}$(\tau, e_w, e_r)$.
		\hfill (\hl{rfto1})
		\item or, $e_w \nin \ordevents{\sc}_\tau$ $\^$
		$\nexists e_w' \in \ordwrites{\sc}_\tau$ \st
		$\hb{\tau}{e_w}{e_w'}$ $\^$ 
		{\tt imm-scr}$(\tau, e_w', e_r)$.
		
		\hfill (\hl{rfto2})
	\end{itemize}
	{\em where}, {\tt imm-scr}$(\tau,a,b)$ $\definedas$ $a \in 
	\ordwrites{\sc}_\tau$, $b \in \ordreads{\sc}_\tau$,
	$\to{\tau}{a}{b}$ and $obj(a) = obj(b)$ $\^$ 
	$\nexists c \in \ordwrites{\sc}_\tau$ \st $obj(c) = obj(a)$ $\^$ 
	$\to{\tau}{a}{\to{\tau}{c}{b}}$.
	\item  $\forall \rf{\tau}{e_w}{e_r}$ \st $e_w \in 
	\ordevents{\sc}_\tau$, $\exists \mathbb{F} \in
	\ordfences{\sc}_\tau$ \st $\seqb{\tau}{\mathbb{F}}{e_r}$
	$\^$ $\to{\tau}{e_w}{\mathbb{F}}$
	$\^$ $\nexists e_w' \in \ordwrites{\sc}$ where
	$\to{\tau}{\to{\tau}{e_w}{e_w'}}{\mathbb{F}}$.
	\hfill (\hl{frfto})
\end{enumerate}

\noindent
We represent the conjunction of the four conditions by (\hl{to-sc});
intuitively,
\begin{itemize}[leftmargin=1em]
\item 
	$\forall e^\sc_1, e^\sc_2 \in \ordevents{\sc}_\tau$ 
	if $\to{\tau}{e^\sc_1}{e^\sc_2}$ then 
	$(e^\sc_2,e^\sc_1)$ $\nin$
	$\setHB$ $\union$ $\setMO$ $\union$ $\setRF$ $\union$ $\setFR$; 
	{\em and},
	
\item 
	an \sc read (or any read with an \sc fence $\setSB$ ordered 
	before it) must not read from an \sc write that is not 
	{\em immediately} $\setTO$ ordered before it. 
	\hfill 
\end{itemize}

Conjunction of  \hlref{coherence conditions} and \hlref{to-sc}
forms the sufficient condition to determine if a trace $\tau$ 
is valid under \cc.

\noindent
{\bf HB with \cc fences.}
\cc fences form $\setITHB$ with other events \cite{batty2011mathematizing,C11}.
A fence can be associated with the memory orders $\rel$, $\acq$, 
$\acqrel$ and $\sc$.
An appropriately placed fence can form $\setSW$ and $\setDOB$ 
relation from an $\setRF$ relation between events of different 
threads, formally:

\begin{figure*}[t]
	\centering
	\begin{tabular}{|c|c|c|c|}
		\hline
		\resizebox{0.18\textwidth}{!}{\tikzset{every picture/.style={line width=0.75pt}} 
\begin{tikzpicture}[x=1em,y=1em,yscale=-1,xscale=-1]
\tikzstyle{every node}=[font=\normalfont]
\node (w1) {$ w^\rel_1 $};
\node (r1) [below=20pt of w1] {$ r^\acq_1 $};

\draw [->,>=stealth,color=Magenta] ($ (w1.south east)+(.3,-5pt) $) to[out=135,in=-135] node[midway,right=-2pt,font=\scriptsize] {\textcolor{black}{\lsw}} ($ (r1.south east)+(0.4,-7pt) $);
\draw [->,>=stealth,color=PineGreen] ($ (w1.south west)+(-.3,-5pt) $) to[out=45,in=-45] node[left=-2pt,pos=.25,font=\scriptsize] {\textcolor{black}{\lrf}}($ (r1.south west)+(-0.3,-7pt) $);

\end{tikzpicture}} &
		\resizebox{0.20\textwidth}{!}{\tikzset{every picture/.style={line width=0.75pt}} 
\begin{tikzpicture}[x=1em,y=1em,yscale=-1,xscale=-1]
\tikzstyle{every node}=[font=\normalfont]
\node (w1) [inner sep=2pt] {$ w^\rel_1 $};
\node (r1) [right=25pt of w1,inner sep=2pt] {$ r_1 $};
\node (f1) [below left=19pt and -15pt of r1,inner sep=2pt] {$ \mathbb{F}^\acq_1 $};

`\draw [->,>=stealth,color=Magenta] (w1.south) -- node[midway,left=0pt,font=\scriptsize,color=black] { $\lsw$ } (f1.west);
\draw [->,>=stealth,color=PineGreen] (w1) -- node[midway,above=-2pt,font=\scriptsize,color=black] { $ \lrf $ }  (r1);
\draw [->,>=stealth,color=CarnationPink] (r1) -- node[midway,left=-2pt,font=\scriptsize,color=black] { $\lsb$ } (f1);

\end{tikzpicture}} &
		\resizebox{0.20\textwidth}{!}{\tikzset{every picture/.style={line width=0.75pt}} 
\begin{tikzpicture}[x=1em,y=1em,yscale=-1,xscale=-1]
\tikzstyle{every node}=[font=\normalfont]
\node [inner sep=2pt] (f1) {$ \mathbb{F}^\rel_1 $};
\node (w1) [below left=21pt and -15pt of f1,inner sep=2pt] {$ w_1 $};
\node (r1) [right=25pt of w1,inner sep=2pt] {$ r^\acq_1 $};

`\draw [->,>=stealth,color=Magenta] (f1.east) -- node[midway,right=0pt,font=\scriptsize,color=black] { $\lsw$ } (r1.north);
\draw [->,>=stealth,color=PineGreen] (w1) -- node[midway,above=-2pt,font=\scriptsize,color=black] { $ \lrf $ } (r1);
\draw [->,>=stealth,color=CarnationPink] (f1) -- node[midway,left=-2pt,font=\scriptsize,color=black] { $\lsb$ } (w1);

\end{tikzpicture}} &
		\resizebox{0.22\textwidth}{!}{\tikzset{every picture/.style={line width=0.75pt}} 
\begin{tikzpicture}[x=1em,y=1em,yscale=-1,xscale=-1]
\tikzstyle{every node}=[font=\normalfont]
\node (f1) [inner sep=2pt] {$ \mathbb{F}^\rel_1 $};
\node (f2) [right=25pt of f1,inner sep=2pt] {$ \mathbb{F}^\acq_2 $};
\node (w1) [below left=20pt and -15pt of f1,inner sep=2pt] {$ w_1 $};
\node (r1) [below left=20pt and -15pt of f2,inner sep=2pt] {$ r_1 $};

`\draw [->,>=stealth,color=Magenta] (f1) -- node[midway,above=-2pt,font=\scriptsize,color=black] { $\lsw$ } (f2);
\draw [->,>=stealth,color=PineGreen] (w1) -- node[midway,above=-2pt,font=\scriptsize,color=black]{\lrf} (r1);
\draw [->,>=stealth,color=CarnationPink] (f1) -- node[midway,left=-2pt,font=\scriptsize,color=black] { $\lsb$ } (w1);
\draw [->,>=stealth,color=CarnationPink] (r1) -- node[midway,left=-2pt,font=\scriptsize,color=black] { $\lsb$ } (f2);


\end{tikzpicture}} \\
		\hline
		\multicolumn{1}{c}{(a)} &
		\multicolumn{1}{c}{(b)} &
		\multicolumn{1}{c}{(c)} &
		\multicolumn{1}{c}{(d)} \\
		
		\cline{2-3}
		\multicolumn{1}{c|}{} &
		\resizebox{0.21\textwidth}{!}{\tikzset{every picture/.style={line width=0.75pt}} 
\begin{tikzpicture}[x=1em,y=1em,yscale=-1,xscale=-1]
\tikzstyle{every node}=[font=\normalfont]
\node (w1) [inner sep=2pt] {$ w^\rel_1 $};
\node (r1) [right=25pt of w1,inner sep=2pt] {$ r^\acq_1 $};
\node (w2) [below left=21pt and -15pt of w1,inner sep=2pt] {$ w_2 $};

`\draw [->,>=stealth,color=Mulberry] (w1.east) -- node[midway,above=-2pt,font=\scriptsize,color=black] { $\ldob$ } (r1.west);
\draw [->,>=stealth,color=PineGreen] (w2) -- node[midway,below=-2pt,font=\scriptsize,color=black] { $ \lrf $ }  (r1);
\draw [->,>=stealth,color=CarnationPink] (w1) -- node[midway,left=-2pt,font=\scriptsize,color=black] { $\lsb$ } (w2);

\end{tikzpicture}} &
		\resizebox{0.22\textwidth}{!}{\tikzset{every picture/.style={line width=0.75pt}} 
\begin{tikzpicture}[x=1em,y=1em,yscale=-1,xscale=-1]
\tikzstyle{every node}=[font=\normalfont]
\node (w1) [inner sep=2pt] {$ w^\rel_1 $};
\node (f1) [right=25pt of w1,inner sep=2pt] {$ \mathbb{F}^\acq_1 $};
\node (w2) [below left=20pt and -15pt of w1,inner sep=2pt] {$ w_2 $};
\node (r1) [below left=20pt and -13pt of f1,inner sep=2pt] {$ r_1 $};

`\draw [->,>=stealth,color=PineGreen] (w2) -- node[midway,above=-2pt,font=\scriptsize,color=black] { $\lrf$ } (r1);
\draw [->,>=stealth,color=CarnationPink] (w1) -- node[midway,left=-2pt,font=\scriptsize,color=black] { $ \lsb $ } (w2);
\draw [->,>=stealth,color=CarnationPink] (r1) -- node[midway,left=-2pt,font=\scriptsize,color=black] { $\lsb$ } (f1);
\draw [->,>=stealth,color=Mulberry] (w1) -- node[midway,above=-2pt,font=\scriptsize,color=black] { $\ldob$ } (f1);

\end{tikzpicture}} &
		\multicolumn{1}{c}{} \\		
		\cline{2-3}
		\multicolumn{1}{c}{} &
		\multicolumn{1}{c}{(e)} &
		\multicolumn{1}{c}{(f)} & 
		\multicolumn{1}{c}{} \\
	\end{tabular}
	\caption{$\setSW$ and $\setDOB$ relations with fences} 
	\label{fig:sw}
\end{figure*}

The $\setSW$ relation is formed with \cc fences as follows:
\newline
$\forall e_1, e_2 \in \events_\tau$ \st $\rf{\tau}{e_1}{e_2}$,
\begin{itemize}
	\item if
	$ord(e_1) \moge \rel$, 
	$\exists \mathbb{F}^\acq \in \fences_\tau$ \st 
	$ord(\mathbb{F}^\acq) \moge \acq$ and 
	$\seqb{\tau}{e_2}{\mathbb{F}^\acq}$ then
	$\sw{\tau}{e_1}{\mathbb{F}^\acq}$;
	
	\item if
	$ord(e_2) \moge \acq$,
	$\exists \mathbb{F}^\rel \in \fences_\tau$ \st
	$ord(\mathbb{F}^\rel) \moge \rel$ and
	$\seqb{\tau}{\mathbb{F}^\rel}{e_1}$ then
	$\sw{\tau}{\mathbb{F}^\rel}{e_2}$;
	
	\item if
	$\exists \mathbb{F}^\rel, \mathbb{F}^\acq \in \fences_\tau$ \st
	$ord(\mathbb{F}^\rel) \moge \rel$, $ord(\mathbb{F}^\acq) \moge \acq$,
	$\seqb{\tau}{\mathbb{F}^\rel}{e_1}$ and
	$\seqb{\tau}{e_2}{\mathbb{F}^\acq}$ then
	$\sw{\tau}{\mathbb{F}^\rel}{\mathbb{F}^\acq}$.
\end{itemize}

\noindent
The conditions described above, leading to a $\setSW$ between 
program events, are diagrammatically  represented in 
Figure~\ref{fig:sw}(a-d).\newline

\noindent
Similarly, the $\setDOB$ relation is formed with \cc fences as follows:
\newline
$\forall e_1, e_2 \in \events_\tau$ \st $\rf{\tau}{e_1}{e_2}$,
if $\exists e_1' \in \writes_\tau$ \st $e_1$ is in {\em release-sequence} of $e_1'$;
and $\exists \mathbb{F}^\acq \in \fences_\tau$ \st $ord(\mathbb{F}^\acq) \moge \acq$ and 
$\seqb{\tau}{e_2}{\mathbb{F}^\acq}$ then
$\dob{\tau}{e_1'}{\mathbb{F}^\acq}$.\newline
The conditions leading to a $\setDOB$ between program events, are 
diagrammatically  represented in Figure~\ref{fig:sw}(e-f).

\section{Invalidating buggy traces with \cc fences} \label{sec:invalidating ce}
%

The key idea behind the proposed techniques is to introduce
fences such that either \hlref{coherence conditions} or \hlref{to-sc}
are violated.
This section introduces two approaches for determining if the
trace is rendered invalid with fences.

Consider $\imm{\tau}$ of a buggy trace $\tau$. 
The candidate fences of $\imm{\tau}$ inflate $\setSB$, $\setSW$, 
$\setDOB$ and $\setITHB$ relations (fences do not 
contribute to $\isetMO$, $\isetRF$ and $\isetFR$). 
The inflated relations are denoted as $\isetSB$, $\isetSW$,
$\isetDOB$ and $\isetITHB$.
We propose {\em \wkfence} and {\em \stfence} to detect the 
invalidity of $\imm{\tau}$. 
%
%

\begin{figure}[t]
	\centering
	\begin{tabular}{rl}
		\fbox{\resizebox{0.48\textwidth}{!}{\tikzset{every picture/.style={line width=0.75pt}} 
\begin{tikzpicture}[x=1em,y=1em,yscale=-1,xscale=-1]
\tikzstyle{every node}=[font=\normalfont]
\node (ix) [inner sep=2pt,color=Brown] {$\mathbb{I}(x,0)$};
\node (iy) [right=30pt of ix, inner sep=2pt,color=Brown] {$\mathbb{I}(y,0)$};
\node (wx) [below left=15pt and -5pt of ix, inner sep=2pt] {$ W^\rlx(x,1)$};
\node (rx1) [right=20pt of wx, inner sep=2pt] {$R^\rlx(x,1)$};
\node (wy) [below=25pt of rx1, inner sep=2pt] {$W^\rlx(y,1)$};
\node (ry) [right=20pt of rx1, inner sep=2pt] {$R^\rlx(y,1)$};
\node (rx0) [below=25pt of ry, inner sep=2pt] {$R^\rlx(x,0)$};
\draw [->,>=stealth,color=RedOrange] (ix) -- node[pos=0.3,left=-1pt,font=\scriptsize,color=black] { $\lmo$ } (wx);
\draw [->,>=stealth,color=RedOrange] (iy.south) to[in=-135,out=90] node[pos=0.1,left=-2pt,font=\scriptsize,color=black] { $\lmo$ } (wy);
\draw [->,>=stealth,color=PineGreen] (wx) -- node[midway,above=-2pt,font=\scriptsize,color=black] { $\lrf$ } (rx1);
\draw [->,>=stealth,color=CarnationPink] (rx1) -- node[midway,left=-2pt,font=\scriptsize,color=black] { $\lsb$ } (wy);
\draw [->,>=stealth,color=PineGreen] (wy) -- node[midway,above=-2pt,font=\scriptsize,color=black] { $\lrf$ } (ry);
\draw [->,>=stealth,color=CarnationPink] (ry) -- node[midway,left=-2pt,font=\scriptsize,color=black] { $\lsb$ } (rx0);

\draw [color=PineGreen] (-10.2,8.3) -- (-10.2,8.7);
\draw [color=PineGreen] (6.5,8.7) -- (-10.2,8.7);
\draw [color=PineGreen] (6.5,0) -- (6.5,8.7);
\draw [->,>=stealth,color=PineGreen] (6.5,0) -- node[midway,below=-2pt,font=\scriptsize,color=black] { $\lrf^{-1}$ } (2,0);

\draw [opacity=0.3] (-0.1,1.5) -- (-0.1,7);
\draw [opacity=0.3] (0.1,1.5) -- (0.1,7);

\draw [opacity=0.3] (-7.3,1.5) -- (-7.3,7);
\draw [opacity=0.3] (-7.5,1.5) -- (-7.5,7);

\end{tikzpicture}}} &
		\fbox{\resizebox{0.48\textwidth}{!}{\tikzset{every picture/.style={line width=0.75pt}} 
\begin{tikzpicture}[x=1em,y=1em,yscale=-1,xscale=-1]
\tikzstyle{every node}=[font=\normalfont]

\node (ix) [inner sep=2pt,color=Brown] {$\mathbb{I}(x,0)$};
\node (iy) [right=30pt of ix, inner sep=2pt,color=Brown] {$\mathbb{I}(y,0)$};
\node (wx) [below left=10pt and -5pt of ix, inner sep=2pt] {$ W^\rlx(x,1)$};
\node (rx1) [right=20pt of wx, inner sep=2pt] {$R^\rlx(x,1)$};
\node (frel) [below=10pt of rx1, inner sep=2pt,color=Blue] {$\mathbb{F}^\acqrel_1$};
\node (wy) [below=10pt of frel, inner sep=2pt] {$W^\rlx(y,1)$};
\node (ry) [right=20pt of rx1, inner sep=2pt] {$R^\rlx(y,1)$};
\node (facq) [below=10pt of ry, inner sep=2pt,color=Blue] {$\mathbb{F}^\acqrel_2$};
\node (rx0) [below=10pt of facq, inner sep=2pt] {$R^\rlx(x,0)$};
\draw [->,>=stealth,color=RedOrange] (ix) -- node[pos=0.3,left=-1pt,font=\scriptsize,color=black] { $\lmo$ } (wx);
\draw [->,>=stealth,dashed,color=RedOrange,opacity=0.7] (iy.south) to[in=-135,out=90] node[pos=0.1,left=-2pt,font=\scriptsize,color=black] { $\lmo$ } (wy);
\draw [->,>=stealth,color=PineGreen] (wx) -- node[midway,above=-2pt,font=\scriptsize,color=black] { $\lrf$ } (rx1);
\draw [->,>=stealth,color=CarnationPink] (rx1) -- node[pos=0.3,left=-2pt,font=\scriptsize,color=black] { $\lsb$ } (frel);
\draw [->,>=stealth,dashed,color=CarnationPink,opacity=0.7] (frel) -- node[pos=0.3,left=-2pt,font=\scriptsize,color=black] { $\lsb$ } (wy);
\draw [->,>=stealth,dashed,color=PineGreen,opacity=0.7] (wy) -- node[pos=0.8,above=-2pt,font=\scriptsize,color=black] { $\lrf$ } (ry);
\draw [->,>=stealth,dashed,color=CarnationPink,opacity=0.7] (ry) -- node[pos=0.3,left=-2pt,font=\scriptsize,color=black] { $\lsb$ } (facq);
\draw [->,>=stealth,color=CarnationPink] (facq) -- node[pos=0.3,left=-2pt,font=\scriptsize,color=black] { $\lsb$ } (rx0);
\draw [->,>=stealth,color=Magenta] (frel) -- node[midway,above=-2pt,font=\scriptsize,color=black] { \lsw } (facq);
\draw [color=PineGreen] (-10.2,8.3) -- (-10.2,8.7);
\draw [color=PineGreen] (6.5,8.7) -- (-10.2,8.7);
\draw [color=PineGreen] (6.5,0) -- (6.5,8.7);
\draw [->,>=stealth,color=PineGreen] (6.5,0) -- node[midway,below=-2pt,font=\scriptsize,color=black] { $\lrf^{-1}$ } (2,0);

\draw [opacity=0.3] (-0.1,2) -- (-0.1,8);
\draw [opacity=0.3] (0.1,2) -- (0.1,8);

\draw [opacity=0.3] (-7.3,2) -- (-7.3,8);
\draw [opacity=0.3] (-7.5,2) -- (-7.5,8);

\end{tikzpicture}}} \\
		
		\multicolumn{1}{c}{\hl{WRIR}} &
		\multicolumn{1}{c}{\hl{WRIR-invalidated}} \\
		
		\fbox{\resizebox{0.33\textwidth}{!}{\tikzset{every picture/.style={line width=0.75pt}} 
\begin{tikzpicture}[x=1em,y=1em,yscale=-1,xscale=-1]
\tikzstyle{every node}=[font=\normalfont]
\node (wx) [inner sep=2pt] {$ W^\sc(x,1) $};
\node (wy) [right=20pt of wx, inner sep=2pt] {$W^\sc(y,1)$};
\node (f1) [below=12pt of wx, inner sep=2pt, color=White] {$\mathbb{F}_{1}^\sc$:};
\node (f2) [below=12pt of wy, inner sep=2pt, color=White] {$\mathbb{F}_{2}^\sc$:};
\node (ry) [below=12pt of f1, inner sep=2pt] {$R^\rlx(y,0)$:};
\node (rx) [below=12pt of f2, inner sep=2pt] {$R^\rlx(x,0)$:};
\draw [->,>=stealth,color=CarnationPink] (wx) -- node[midway,right=-2pt,font=\scriptsize,color=black] { $\lsb$ } (ry);
\draw [->,>=stealth,color=CarnationPink] (wy) -- node[midway,left=-2pt,font=\scriptsize,color=black] { $\lsb$ } (rx);

\draw [opacity=0.3] (-3.4,-0.5) -- (-3.4,6);
\draw [opacity=0.3] (-3.6,-0.5) -- (-3.6,6);

\end{tikzpicture}}} &
		\fbox{\resizebox{0.33\textwidth}{!}{\tikzset{every picture/.style={line width=0.75pt}} 
\begin{tikzpicture}[x=1em,y=1em,yscale=-1,xscale=-1]
\tikzstyle{every node}=[font=\normalfont]
\node (wx) [inner sep=2pt] {$ W^\sc(x,1) $};
\node (wy) [right=20pt of wx, inner sep=2pt] {$W^\sc(y,1)$};
\node (f1) [below=12pt of wx, inner sep=2pt, color=Blue] {$\mathbb{F}_{1}^\sc$:};
\node (f2) [below=12pt of wy, inner sep=2pt, color=Blue] {$\mathbb{F}_{2}^\sc$:};
\node (ry) [below=12pt of f1, inner sep=2pt] {$R^\rlx(y,0)$:};
\node (rx) [below=12pt of f2, inner sep=2pt] {$R^\rlx(x,0)$:};
\draw [->,>=stealth,color=Mahogany] (wx) -- node[midway,right=-2pt,font=\scriptsize,color=black] { $\lso$ } (f1);
\draw [->,>=stealth,color=Mahogany] (wy) -- node[midway,left=-2pt,font=\scriptsize,color=black] { $\lso$ } (f2);
\draw [->,>=stealth,color=Mahogany] (f1) -- node[pos=0.2,below=-2pt,font=\scriptsize,color=black] { $\lso$ } (wy);
\draw [->,>=stealth,color=Mahogany] (f2) -- node[pos=0.2,below=-2pt,font=\scriptsize,color=black] { $\lso$ } (wx);

\draw [->,>=stealth,dashed,opacity=0.5,color=RoyalPurple] (ry) -- node[pos=0.2,left=-1pt,font=\scriptsize,color=black] { $\lfr$ } (wy);
\draw [->,>=stealth,dashed,opacity=0.5,color=RoyalPurple] (rx) -- node[pos=0.2,right=-1pt,font=\scriptsize,color=black] { $\lfr$ } (wx);

\draw [opacity=0.3] (-3.4,-0.5) -- (-3.4,6);
\draw [opacity=0.3] (-3.6,-0.5) -- (-3.6,6);

\end{tikzpicture}}} \\
				
		\multicolumn{1}{c}{\tab\tab\tab\hl{SB}} &
		\multicolumn{1}{c}{\hl{SB-inv}\tab\tab\tab} \\
	\end{tabular}
\end{figure}

\noindent
{\bf \wkfence.}
\wkfence computes compositions of relations that correspond to the
\hlref{coherence conditions}. It then checks if there
exist cycles in the compositions (using {\em Johnson's algorithm}
\cite{johnson1975finding}).
The approach assumes the memory
order \acqrel for all candidate fences.
Consider a buggy trace \hlref{WRIR} where $x$ and $y$ have $0$ as
initial values.
\wkfence detects a cycle in $\setMO;\setRF;\setHB;\setRF^{-1}$
with the addition of candidate fences $\mathbb{F}^\acqrel_{1}$  and 
$\mathbb{F}^\acqrel_{2}$ as shown in \hlref{WRIR-invalidated}. This 
violates the condition \hlref{co-mrhi}, thus, invalidating \hlref{WRIR}.
Lemma~\ref{lem:weak-sound} shows the correctness of \wkfence. 


\noindent
{\bf \stfence.} This technique works with the assumption that all candidate fences
have the order \sc.
\stfence detects the infeasibility in constructing a $\isetTO$ that 
 adheres to \hlref{to-sc}.
 In order to detect violation
 of \hlref{to-sc}, \stfence introduces a possibly reflexive relation
 on \sc-ordered events of $\imm{\tau}$, called \sc{\em -order}
 $(\isetSO)$.
The $\isetSO$ relation is such that a total
order cannot be formed on the \sc events of $\imm{\tau}$ iff a cycle
exists in $\isetSO$.
%
All \sc event pairs ordered by $\isetHB$, $\isetMO$, $\isetRF$ 
and $\isetFR$ are contained in $\isetSO$.
%
%
Notably, pairs of \sc events that do not have a definite order are
not ordered by $\isetSO$. This is because if such a pair of events is 
involved in a cycle then we can freely flip their order and 
eliminate the cycle.
Consider the buggy trace \hlref{SB}, 
$\to{}{W^\sc(x,1)}{W^\sc(y,1)}$ and $\to{}{W^\sc(y,1)}{W^\sc(x,1)}$
are both valid total-orders on the \sc events of the trace.
The set $\isetSO$ does not contain either of the two event pairs and 
would be empty for this example.
%

As a consequence, pairs of events that do not have definite total
order cannot contribute to the reflexivity of $\setSO$ and can be
safely ignored.  
Thus, $\setSO^+ \subseteq \setTO$ for a trace $\tau$.
Further, if a total order cannot be formed
on \sc ordered events then a corresponding cycle exists in $\isetSO$.
The observations are formally presented with supporting proofs in 
Lemmas~\ref{lem:so subset to},\ref{lem:strong-sound}.
Lemma~\ref{lem:strong-sound} further proves that \stfence is sound.

\noindent
Definition~\ref{def:so} formally presents $\isetSO$
based on the above stated considerations.
\begin{definition}{\bf \sc-order ($\so{\imm{\tau}}{}{}$)}\newline
	\normalfont
%
	$\forall e_1, e_2 \in \events_\tau$ \st 
	$(e_1,e_2) \in$ $R$, where $R$ = $\setHB$ $\union$ $\setMO$ $\union$ $\setRF$ 
	$\union$ $\setFR$
%
\begin{itemize}
	\item if
	$e_1, e_2 \in \ordevents{\sc}_\tau$ then 
	$\so{\imm{\tau}}{e_1}{e_2}$;	
	\hfill(\hl{soee})
		
	\item if
	$e_1 \in \ordevents{\sc}_\tau$, 
	$\exists \mathbb{F}^\sc \in \ordfences{\sc}_{\imm{\tau}}$ \st
	$\seqb{\imm{\tau}}{e_2}{\mathbb{F}^\sc}$ then
	$\so{\imm{\tau}}{e_1}{\mathbb{F}^\sc}$;
	\hfill(\hl{soef})
	
	\item if
	$e_2 \in \ordevents{\sc}_\tau$, 
	$\exists \mathbb{F}^\sc \in \ordfences{\sc}_{\imm{\tau}}$ \st
	$\seqb{\imm{\tau}}{\mathbb{F}^\sc}{e_1}$ then
	$\so{\imm{\tau}}{\mathbb{F}^\sc}{e_2}$;
	\hfill(\hl{sofe})
	
	\item if
	$\exists \mathbb{F}^\sc_1$, $\mathbb{F}^\sc_2$ 
	$\in \ordfences{\sc}_{\imm{\tau}}$ \st
	$\seqb{\imm{\tau}}{\mathbb{F}^\sc_1}{e_1}$ and 
	$\seqb{\imm{\tau}}{e_2}{\mathbb{F}^\sc_2}$ then
	$\so{\imm{\tau}}{\mathbb{F}^\sc_1}{\mathbb{F}^\sc_2}$.
	\hfill(\hl{soff})
\end{itemize}
\label{def:so}
\end{definition}
The trace depicted in \hlref{SB} can be invalidated with
strong fences as shown in \hlref{SB-inv}. The \sc events of
\hlref{SB-inv} cannot be totally ordered
 and \stfence detects the same through a cycle
in $\so{}{}{}$ (formed by \hlref{soee} and \hlref{sofe}).

\begin{figure*}[t]
\centering       
\begin{tabular}{|c|c|c|c|}
		\hline
		\resizebox{0.17\textwidth}{!}{\tikzset{every picture/.style={line width=0.75pt}} 
\begin{tikzpicture}[x=1em,y=1em,yscale=-1,xscale=-1]
\tikzstyle{every node}=[font=\normalfont]
\node (w1) {$ e^\sc_1 $};
\node (r1) [below=15pt of w1] {$ e^\sc_2 $};

\draw [->,>=stealth,color=Mahogany] ($ (w1.south east)+(.3,-5pt) $) to[out=135,in=-135] node[midway,right=-2pt,font=\scriptsize] {\textcolor{black}{\lso}} ($ (r1.south east)+(0.4,-7pt) $);
\draw [->,>=stealth,color=Cerulean] ($ (w1.south west)+(-.3,-5pt) $) to[out=45,in=-45] node[left=-2pt,midway,font=\scriptsize] {$R$}($ (r1.south west)+(-0.3,-7pt) $);

\end{tikzpicture}} &
		\resizebox{0.20\textwidth}{!}{\tikzset{every picture/.style={line width=0.75pt}} 
\begin{tikzpicture}[x=1em,y=1em,yscale=-1,xscale=-1]
\tikzstyle{every node}=[font=\normalfont]
\node (w1) [inner sep=2pt] {$ e^\sc_1 $};
\node (r1) [right=25pt of w1,inner sep=2pt] {$ e_2 $};
\node (f1) [below left=19pt and -15pt of r1,inner sep=2pt] {$ \mathbb{F}^\sc $};

`\draw [->,>=stealth,color=Mahogany] (w1.south) -- node[midway,left=0pt,font=\scriptsize,color=black] { $\lso$ } (f1.west);
\draw [->,>=stealth,color=Cerulean] (w1) -- node[midway,above=-2pt,font=\scriptsize] { $R$ }  (r1);
\draw [->,>=stealth,color=CarnationPink] (r1) -- node[midway,left=-2pt,font=\scriptsize,color=black] { $\lsb$ } (f1);

\end{tikzpicture}} &
        \resizebox{0.21\textwidth}{!}{\tikzset{every picture/.style={line width=0.75pt}} 
\begin{tikzpicture}[x=1em,y=1em,yscale=-1,xscale=-1]
\tikzstyle{every node}=[font=\normalfont]
\node [inner sep=2pt] (f1) {$ \mathbb{F}^\sc $};
\node (w1) [below left=21pt and -15pt of f1,inner sep=2pt] {$ e_1 $};
\node (r1) [right=25pt of w1,inner sep=2pt] {$ e^\sc_2 $};

`\draw [->,>=stealth,color=Mahogany] (f1.east) -- node[midway,right=0pt,font=\scriptsize,color=black] { $\lso$ } (r1.north);
\draw [->,>=stealth,color=Cerulean] (w1) -- node[midway,above=-2pt,font=\scriptsize] { $R$ } (r1);
\draw [->,>=stealth,color=CarnationPink] (f1) -- node[midway,left=-2pt,font=\scriptsize,color=black] { $\lsb$ } (w1);

\end{tikzpicture}} &                                    
        \resizebox{0.22\textwidth}{!}{\tikzset{every picture/.style={line width=0.75pt}} 
\begin{tikzpicture}[x=1em,y=1em,yscale=-1,xscale=-1]
\tikzstyle{every node}=[font=\normalfont]
\node (f1) [inner sep=2pt] {$ \mathbb{F}^\sc_1 $};
\node (f2) [right=25pt of f1,inner sep=2pt] {$ \mathbb{F}^\sc_2 $};
\node (w1) [below left=20pt and -15pt of f1,inner sep=2pt] {$ e_1 $};
\node (r1) [below left=20pt and -15pt of f2,inner sep=2pt] {$ e_2 $};

`\draw [->,>=stealth,color=Mahogany] (f1) -- node[midway,above=-2pt,font=\scriptsize,color=black] { $\lso$ } (f2);
\draw [->,>=stealth,color=Cerulean] (w1) -- node[midway,above=-2pt,font=\scriptsize] {$R$} (r1);
\draw [->,>=stealth,color=CarnationPink] (f1) -- node[midway,left=-2pt,font=\scriptsize,color=black] { $\lsb$ } (w1);
\draw [->,>=stealth,color=CarnationPink] (r1) -- node[midway,left=-2pt,font=\scriptsize,color=black] { $\lsb$ } (f2);


\end{tikzpicture}} \\      
                \hline
		\multicolumn{1}{c}{\hlref{soee}} &
		\multicolumn{1}{c}{\hlref{soef}} &
		\multicolumn{1}{c}{\hlref{sofe}} &
		\multicolumn{1}{c}{\hlref{soff}} \\
	\end{tabular}
\end{figure*}

\noindent
{\bf Scope of \ourtechnique/\fasttechnique.}
\begin{figure}[t]
	\centering
	\begin{minipage}{\textwidth}
		\setlength{\tabcolsep}{0pt}
		\begin{tabular}{|c|c|}
			\hline
			\resizebox{0.48\textwidth}{!}{\tikzset{every picture/.style={line width=0.75pt}} 
\begin{tikzpicture}[x=1em,y=1em,yscale=1,xscale=1]
\tikzstyle{every node}=[font=\small]

\node (init) {$Initially$, $x=0$, $y=0$};
\node (f11) [below left=0pt and 20pt of init,color=White] { $F^\sc_{11}$ };
\node (wx1) [below=8pt of f11] {$ W^\sc(x,1) $};
\node (f12) [below=8pt of wx1,color=White] { $F^\sc_{12}$ };

\node (f21) [right=25pt of f11,color=White] { $F^\sc_{21}$ };
\node (rx1) [below=8pt of f21] {$R^\sc(x,1) $};
\node (f22) [below=8pt of rx1,color=White] { $F^\sc_{22}$ };
\node (ry0) [below=8pt of f22] {$R^\sc(y,0) $};
\node (f23) [below=8pt of ry0,color=White] { $F^\sc_{23}$ };

\node (f31) [right=25pt of f21,color=White] { $F^\sc_{31}$ };
\node (wy1) [below=8pt of f31] {$ W^\sc(y,1) $};
\node (f32) [below=8pt of wy1,color=White] { $F^\sc_{32}$ };

\node (f41) [right=25pt of f31,color=White] { $F^\sc_{41}$ };
\node (ry1) [below=8pt of f41] {$R^\sc(y,1) $};
\node (f42) [below=8pt of ry1,color=White] { $F^\sc_{42}$ };
\node (rx0) [below=8pt of f42] {$R^\sc(x,0) $};
\node (f43) [below=8pt of rx0,color=White] { $F^\sc_{43}$ };

\draw [->,>=stealth,color=Mahogany] (rx1) -- (ry0);
\draw [->,>=stealth,color=Mahogany] (ry1) -- (rx0);

\draw [->,>=stealth,color=Mahogany] ($ (wx1.east)+(-3pt,0) $) -- ($ (rx1.west)+(3pt,0) $);
\draw [->,>=stealth,color=Mahogany] (ry0) -- (wy1);
\draw [->,>=stealth,color=Mahogany] ($ (wy1.east)+(-3pt,0) $) -- ($ (ry1.west)+(3pt,0) $);
\draw [color=Mahogany] (rx0.south)+(0,3pt) -- ($ (rx0.south)+(0,-13.2pt) $);
\draw [color=Mahogany] ($ (rx0.south)+(0,-13.2pt) $) -- ($ (wx1.south)+(0,-60pt) $);
\draw [->,>=stealth,color=Mahogany] ($ (wx1.south)+(0,-60pt) $) -- (wx1);

\draw [gray,opacity=0.5] (-6.2,-2.5) -- (-6.2,-12.5);
\draw [gray,opacity=0.5] (-6.0,-2.5) -- (-6.0,-12.5);

\draw [gray,opacity=0.5] (-1.3,-2.5) -- (-1.3,-12.5);
\draw [gray,opacity=0.5] (-1.1,-2.5) -- (-1.1,-12.5);

\draw [gray,opacity=0.5] (3.7,-2.5) -- (3.7,-12.5);
\draw [gray,opacity=0.5] (3.9,-2.5) -- (3.9,-12.5);

\end{tikzpicture}} &
			\resizebox{0.5\textwidth}{!}{\tikzset{every picture/.style={line width=0.75pt}} 
\begin{tikzpicture}[x=1em,y=1em,yscale=1,xscale=1]
\tikzstyle{every node}=[font=\small]

\node (init) {$Initially$, $x=0$, $y=0$};
\node (f11) [below left=0pt and 20pt of init,color=Blue] { $\mathbb{F}^\sc_{11}$ };
\node (wx1) [below=8pt of f11] {$ W^\rlx(x,1) $};
\node (f12) [below=8pt of wx1,color=Blue] { $\mathbb{F}^\sc_{12}$ };

\node (f21) [right=25pt of f11,color=Blue] { $\mathbb{F}^\sc_{21}$ };
\node (rx1) [below=8pt of f21] {$R^\rlx(x,1) $};
\node (f22) [below=8pt of rx1,color=Blue] { $\mathbb{F}^\sc_{22}$ };
\node (ry0) [below=8pt of f22] {$R^\rlx(y,0) $};
\node (f23) [below=8pt of ry0,color=Blue] { $\mathbb{F}^\sc_{23}$ };

\node (f31) [right=25pt of f21,color=Blue] { $\mathbb{F}^\sc_{31}$ };
\node (wy1) [below=8pt of f31] {$ W^\rlx(y,1) $};
\node (f32) [below=8pt of wy1,color=Blue] { $\mathbb{F}^\sc_{32}$ };

\node (f41) [right=25pt of f31,color=Blue] { $\mathbb{F}^\sc_{41}$ };
\node (ry1) [below=8pt of f41] {$R^\rlx(y,1) $};
\node (f42) [below=8pt of ry1,color=Blue] { $\mathbb{F}^\sc_{42}$ };
\node (rx0) [below=8pt of f42] {$R^\rlx(x,0) $};
\node (f43) [below=8pt of rx0,color=Blue] { $\mathbb{F}^\sc_{43}$ };

\draw [->,>=stealth,color=Mahogany] (f11) -- (wx1);
\draw [->,>=stealth,color=Mahogany] (wx1) -- (f12);

\draw [->,>=stealth,color=Mahogany] (f21) -- (rx1);
\draw [->,>=stealth,color=Mahogany] (rx1) -- (f22);
\draw [->,>=stealth,color=Mahogany] (f22) -- (ry0);
\draw [->,>=stealth,color=Mahogany] (ry0) -- (f23);

\draw [->,>=stealth,color=Mahogany] (f31) -- (wy1);
\draw [->,>=stealth,color=Mahogany] (wy1) -- (f32);

\draw [->,>=stealth,color=Mahogany] (f41) -- (ry1);
\draw [->,>=stealth,color=Mahogany] (ry1) -- (f42);
\draw [->,>=stealth,color=Mahogany] (f42) -- (rx0);
\draw [->,>=stealth,color=Mahogany] (rx0) -- (f43);

\draw [->,>=stealth,color=Mahogany] (f11) -- (f22);
\draw [->,>=stealth,color=Mahogany] (f22) -- (f32);
\draw [->,>=stealth,color=Mahogany] (f31) -- (f42);
\draw [color=Mahogany] (f42.east)+(0,0) -- ($ (f42.east)+(20pt,0) $);
\draw [color=Mahogany] (f42.east)+(20pt,0) -- ($ (f43.east)+(20pt,-7pt) $);
\draw [color=Mahogany] ($ (f43.east)+(20pt,-7pt) $) -- ($ (f12.south)+(-0.05,-47.2pt) $);
\draw [->,>=stealth,color=Mahogany] ($ (f12.south)+(0,-47.2pt) $) -- (f12);

\draw [gray,opacity=0.5] (-6.2,-1.0) -- (-6.2,-12.5);
\draw [gray,opacity=0.5] (-6.0,-1.0) -- (-6.0,-12.5);

\draw [gray,opacity=0.5] (-1.3,-1.0) -- (-1.3,-12.5);
\draw [gray,opacity=0.5] (-1.1,-1.0) -- (-1.1,-12.5);

\draw [gray,opacity=0.5] (3.7,-1.0) -- (3.7,-12.5);
\draw [gray,opacity=0.5] (3.9,-1.0) -- (3.9,-12.5);

\end{tikzpicture}} \\
			\hline
			\multicolumn{1}{c}{\hl{iriw-invalid}} & 
			\multicolumn{1}{c}{\hl{iriw-valid}} \\
		\end{tabular}
	\end{minipage}
\end{figure}

\begin{figure}[t]
	\centering
	\begin{minipage}{\textwidth}
		\setlength{\tabcolsep}{0pt}
		\begin{tabular}{|c|c||c|c|}
			\multicolumn{2}{c}{
				\fbox{
					\resizebox{0.3\textwidth}{!}{
						\tikzset{every picture/.style={line width=0.75pt}} 
\begin{tikzpicture}[x=1em,y=1em,yscale=1,xscale=1]
\tikzstyle{every node}=[font=\small]

\node (init) {$Initially$, $x=0$, $y=0$};
\node (wx) [below left=0pt and -40pt of init] {$ W^\sc(x,1) $};
\node (f12) [below=1pt of wx,color=White] { $\mathbb{F}^\sc_{1}$ };
\node (ry) [below=1pt of f12] {$ R^\sc(y,0) $};

\node (wy) [right=25pt of wx] {$ W^\sc(y,1) $};
\node (f22) [below=1pt of wy,color=White] { $\mathbb{F}^\sc_{2}$ };
\node (rx) [below=1pt of f22] {$ R^\sc(x,0) $};

\draw [->,>=stealth,color=Mahogany] (wx) -- (ry);
\draw [->,>=stealth,color=Mahogany] (ry) -- (wy);
\draw [->,>=stealth,color=Mahogany] (wy) -- (rx);
\draw [->,>=stealth,color=Mahogany] (rx) -- (wx);

\draw [gray,opacity=0.5] (0.2,-1) -- (0.2,-6);
\draw [gray,opacity=0.5] (0.4,-1) -- (0.4,-6);

\end{tikzpicture}}}} &
			\multicolumn{2}{c}{
				\fbox{
					\resizebox{0.3\textwidth}{!}{
						\tikzset{every picture/.style={line width=0.75pt}} 
\begin{tikzpicture}[x=1em,y=1em,yscale=1,xscale=1]
\tikzstyle{every node}=[font=\small]

\node (init) {$Initially$, $x=0$, $y=0$};
\node (wx) [below left=0pt and -40pt of init] {$ W^\rlx(x,1) $};
\node (f12) [below=1pt of wx,color=Blue] { $\mathbb{F}^\sc_{1}$ };
\node (ry) [below=1pt of f12] {$ R^\rlx(y,0) $};

\node (wy) [right=25pt of wx] {$ W^\rlx(y,1) $};
\node (f22) [below=1pt of wy,color=Blue] { $\mathbb{F}^\sc_{2}$ };
\node (rx) [below=1pt of f22] {$ R^\rlx(x,0) $};

\draw [->,>=stealth,color=Mahogany] ($ (f12.north east)+(-2pt,-3pt) $) to[out=25,in=155] ($ (f22.north west)+(0pt,-3pt) $);
\draw [->,>=stealth,color=Mahogany] ($ (f22.south west)+(0pt,5pt) $) to[out=-155,in=-25] ($ (f12.south east)+(-2pt,5pt) $);

\draw [gray,opacity=0.5] (0.2,-1) -- (0.2,-6);
\draw [gray,opacity=0.5] (0.4,-1) -- (0.4,-6);

\end{tikzpicture}}}} \\
			\multicolumn{2}{c}{\hl{SB-inv-mo}} & 
			\multicolumn{2}{c}{\hl{SB-inv-fen}} \\
			
			\hline
			\resizebox{0.25\textwidth}{!}{\tikzset{every picture/.style={line width=0.75pt}} 
\begin{tikzpicture}[x=1em,y=1em,yscale=1,xscale=1]
\tikzstyle{every node}=[font=\footnotesize]

\node (init) {$Initially$, $x=0$, $y=0$};
\node (f11) [below left=0pt and -40pt of init,color=Brown] {\tt DMB ish };
\node (wx) [below=1pt of f11] {$ str(x,1) $};
\node (f12) [below=1pt of wx,color=Brown] {\tt DMB ish  };
\node (ry) [below=1pt of f12] {$ ld(y) $};
\node (f13) [below=1pt of ry,color=Brown] {\tt DMB ish  };

\node (f21) [right=25pt of f11,color=Brown] {\tt DMB ish };
\node (wy) [below=1pt of f21] {$ str(y,1) $};
\node (f22) [below=1pt of wy,color=Brown] {\tt DMB ish  };
\node (rx) [below=1pt of f22] {$ ld(x) $};
\node (f23) [below=1pt of rx,color=Brown] {\tt DMB ish  };

\draw [gray,opacity=0.5] (0.2,-1) -- (0.2,-8.5);
\draw [gray,opacity=0.5] (0.4,-1) -- (0.4,-8.5);

\end{tikzpicture}} &
			\resizebox{0.24\textwidth}{!}{\tikzset{every picture/.style={line width=0.75pt}} 
\begin{tikzpicture}[x=1em,y=1em,yscale=1,xscale=1]
\tikzstyle{every node}=[font=\footnotesize]

\node (init) {$Initially$, $x=0$, $y=0$};
\node (f11) [below left=0pt and -40pt of init,color=Brown] {\tt hwsync };
\node (wx) [below=1pt of f11] {$ str(x,1) $};
\node (f12) [below=1pt of wx,color=Brown] {\tt hwsync  };
\node (ry) [below=1pt of f12] {$ ld(y) $};
\node (f13) [below=1pt of ry,color=Brown] {\tt isync+ };

\node (f21) [right=25pt of f11,color=Brown] {\tt hwsync };
\node (wy) [below=1pt of f21] {$ str(y,1) $};
\node (f22) [below=1pt of wy,color=Brown] {\tt hwsync  };
\node (rx) [below=1pt of f22] {$ ld(x) $};
\node (f23) [below=1pt of rx,color=Brown] {\tt isync+  };

\draw [gray,opacity=0.5] (0.2,-1) -- (0.2,-8.5);
\draw [gray,opacity=0.5] (0.4,-1) -- (0.4,-8.5);

\end{tikzpicture}} &
			\resizebox{0.25\textwidth}{!}{\tikzset{every picture/.style={line width=0.75pt}} 
\begin{tikzpicture}[x=1em,y=1em,yscale=1,xscale=1]
\tikzstyle{every node}=[font=\footnotesize]

\node (init) {$Initially$, $x=0$, $y=0$};
\node (f11) [below left=0pt and -40pt of init,color=White] {\tt DMB ish };
\node (wx) [below=1pt of f11] {$ str(x,1) $};
\node (f12) [below=1pt of wx,color=Brown] {\tt DMB ish  };
\node (ry) [below=1pt of f12] {$ ld(y) $};
\node (f13) [below=1pt of ry,color=White] {\tt DMB ish  };

\node (f21) [right=25pt of f11,color=White] {\tt DMB ish };
\node (wy) [below=1pt of f21] {$ str(y,1) $};
\node (f22) [below=1pt of wy,color=Brown] {\tt DMB ish  };
\node (rx) [below=1pt of f22] {$ ld(x) $};
\node (f23) [below=1pt of rx,color=White] {\tt DMB ish  };

\draw [gray,opacity=0.5] (0.2,-1) -- (0.2,-8.5);
\draw [gray,opacity=0.5] (0.4,-1) -- (0.4,-8.5);

\end{tikzpicture}} &
			\resizebox{0.25\textwidth}{!}{\tikzset{every picture/.style={line width=0.75pt}} 
\begin{tikzpicture}[x=1em,y=1em,yscale=1,xscale=1]
\tikzstyle{every node}=[font=\footnotesize]

\node (init) {$Initially$, $x=0$, $y=0$};
\node (f11) [below left=0pt and -40pt of init,color=White] {\tt DMB ish };
\node (wx) [below=1pt of f11] {$ str(x,1) $};
\node (f12) [below=1pt of wx,color=Brown] {\tt hwsync  };
\node (ry) [below=1pt of f12] {$ ld(y) $};
\node (f13) [below=1pt of ry,color=White] {\tt DMB ish  };

\node (f21) [right=25pt of f11,color=White] {\tt DMB ish };
\node (wy) [below=1pt of f21] {$ str(y,1) $};
\node (f22) [below=1pt of wy,color=Brown] {\tt hwsync  };
\node (rx) [below=1pt of f22] {$ ld(x) $};
\node (f23) [below=1pt of rx,color=White] {\tt DMB ish  };

\draw [gray,opacity=0.5] (0.2,-1) -- (0.2,-8.5);
\draw [gray,opacity=0.5] (0.4,-1) -- (0.4,-8.5);

\end{tikzpicture}} \\
			\hline
			
			\multicolumn{1}{c}{\hl{SB-inv-mo-ARM}} &
			\multicolumn{1}{c}{\hl{SB-inv-mo-power}\footnotemark} &
			\multicolumn{1}{c}{\hl{SB-inv-fen-ARM}} &
			\multicolumn{1}{c}{\hl{SB-inv-fen-power}} \\
		\end{tabular}
	\end{minipage}
\end{figure}
\footnotetext{{\tt isync+} refers to {\tt cmp; bc; isync}}

This work synthesizes \cc fences and stands fundamentally
different from techniques that modify the memory orders of
program events.
\sc \cc fences cannot restore sequential consistency~\cite{LahavVafeiadis-PLDI17},
hence, strengthening memory orders may invalidate buggy traces 
that the strongest \cc fences cannot.

The examples \hlref{iriw-invalid} and \hlref{iriw-valid} 
highlight the difference, where \textcolor{Mahogany}{\bf $\rightarrow$}
depicts the total-order on \sc events.
The fences of \hlref{iriw-valid} do not introduce an ordering
between the write events and the corresponding read events
thereby allowing reads from initial events.
Thus, the trace shown in \hlref{iriw-valid} cannot be invalidated by 
\cc fences. 
However, changing the memory order of read and write events
achieve the desired ordering and invalidate the outcome, 
as shown in \hlref{iriw-invalid}.

\ourtechnique and \fasttechnique invalidate traces by synthesizing 
\cc fences, thus, \hlref{iriw-valid} cannot be stopped by \ourtechnique 
or \fasttechnique.
However, architectures translate the strong memory access events 
to memory access operation and supporting barriers.
The translation of strongly ordered event to barriers may be
sub-optimal.

The examples \hlref{SB-inv-mo} and \hlref{SB-inv-fen} 
highlight the difference. The trace shown in the examples is
invalidated by strengthening the memory orders of writes and
reads in \hlref{SB-inv-mo} and by synthesizing \sc fences in
\hlref{SB-inv-fen}.
The translation of \hlref{SB-inv-mo} to Power and ARMv7 is 
depicted in \hlref{SB-inv-mo-power} and \hlref{SB-inv-mo-ARM}
respectively.
The translation of \hlref{SB-inv-fen} to Power and ARMv7 is 
depicted in \hlref{SB-inv-fen-power} and \hlref{SB-inv-fen-ARM}
respectively.
Clearly each of the two architectures place additional (and
unnecessary) barriers on interpreting barrier requirement
from memory orders.

\section{Methodology} \label{sec:methodology}
\noindent
{\bf Buggy traces and candidate fences.}
Algorithms~\ref{alg:main algo} and~\ref{alg:fast algo} present \ourtechnique 
and  \fasttechnique, respectively. 
The algorithms rely on an external buggy trace generator (\btg)
for the buggy trace(s) of $P$ (lines 2,8).
The candidate fences are inserted (to obtain
$\imm{\tau}$), and the event relations are updated (lines 16-17).

\noindent
{\bf Detecting violation of trace coherence.}
The algorithms detect possible violations of trace coherence
conditions resulting from the candidate fences at lines 18-19
of the function {\tt synthesisCore}.
Figures \hlref{RWRW-inv-sync} and \hlref{RWRW-inv-sync-opt} 
represent two instances of violations of \hlref{co-rh} detected by
\wkfence (through a cycle 
$W^\rlx(y,1){\isetRF}R^\rlx(y,1){\isetHB}W^\rlx(y,1)$). 
The candidate solutions corresponding to these cycles (which include 
only candidate fences) are $\{\mathbb{F}_{12},\mathbb{F}_{22}\}$ and 
$\{\mathbb{F}_{12}\}$. Further, for the same example, 
\hlref{RWRW-inv-to} represents a violation detected by \stfence with 
the candidate solution $\{\mathbb{F}_{12},\mathbb{F}_{22}\}$. 
The algorithms discard all candidate fences other than $\mathbb{F}_{12}$ 
and $\mathbb{F}_{22}$ from future considerations (assuming no other 
violations were detected). Now $\tau$ can be invalidated as the set of 
cycles is nonempty (line 20).

The complexity of detecting all cycles for a trace is
$\bigo$(($|\events_\tau|$+{\tt E}).({\tt C}+1)) where {\tt C}
represents the number of cycles of $\tau$ and {\tt E} represents the
number of pairs of events in $\events_\tau$.  Note that {\tt E} is in
$O(|\events_\tau|^2)$ and {\tt C} is in $O(|\events_\tau|!)$. Thus,
Weak- and Strong-Fensying have exponential complexities in
the number of traces and the number of events per trace.

\begin{figure}[t]
\begin{minipage}{0.48\textwidth}
	\input{algos/fensying.tex}
\end{minipage}
\hfill
\begin{minipage}{0.48\textwidth}
	\input{algos/fastfensying.tex}
\end{minipage}

\input{algos/fensyingcore.tex}
\end{figure}

\noindent
{\bf Reduction for optimality.}
The algorithms use a \sat solver to determine the optimal  
number of candidate fences.
The candidate fences from each candidate solution of $\tau$ are
conjuncted to form a \sat query.  Further, to retain at least one
solution corresponding to $\tau$ the algorithms take a disjunction 
of the conjuncts.  The \sat query is represented in
the algorithm as $\Phi_\tau := \formula{{\tt weakCycles}_\tau \v {\tt
strongCycles}_\tau}$ (line 22) and presented in Equation~\ref{eq:sat
query trace} (where ${\tt \bf W_\tau}$ and ${\tt \bf S_\tau}$
represent ${\tt weakCycles}_\tau$ and ${\tt strongCycles}_\tau$ and
${\tt W}^\mathbb{F}$ and ${\tt S}^\mathbb{F}$ represent the set of
candidate fences in cycles {\tt W} and {\tt S} respectively).
Further, \ourtechnique combines the \sat formulas corresponding to each 
buggy trace via conjunction  (line 4), shown in Equation~\ref{eq:sat query}.
However, note that for \fasttechnique $\Phi = \Phi_\tau$.
\newline
\begin{minipage}[b][][b]{0.6\linewidth}
	\begin{equation}\label{eq:sat query trace}
		\Phi_\tau = 
		(\bigvee\limits_{{\tt W} \in {\tt \bf W_\tau}}
		\bigwedge\limits_{\mathbb{F}_{\tt w} \in {\tt W}^\mathbb{F}  } 
		\mathbb{F}_{\tt w})
		\v
		(\bigvee\limits_{{\tt S} \in {\tt \bf S_\tau}}
		\bigwedge\limits_{\mathbb{F}_{\tt s} \in {\tt S}^\mathbb{F}} 
		\mathbb{F}_{\tt s})
	\end{equation}
\end{minipage}%
\begin{minipage}[b][][b]{0.4\linewidth}
	\begin{equation}\label{eq:sat query}
		\Phi = \bigwedge\limits_{\tau \in {\tt BT}} \Phi_\tau
	\end{equation}
\end{minipage}
%
\ \\ \ \\
We use a \sat solver to compute the {\em min-model} 
({\tt min}$\Phi$) of the query $\Phi$ (lines 5,10).
%
For instance, the query for  \hlref{RWRW-bt} is $\Phi$ =
($\mathbb{F}_{12}$) $\v$ ($\mathbb{F}_{12}$ $\^$ $\mathbb{F}_{22}$) 
$\v$ ($\mathbb{F}_{12}$ $\^$ $\mathbb{F}_{22}$)
and min-model, {\tt min}$\Phi$ = $\{\mathbb{F}_{12}\}$.
%
%
The solution to the \sat query returns the smallest set of fences to
be synthesized.

The complexity of constructing the query $\Phi_\tau$ is $\bigo$({\tt
C}.{\tt F}), where {\tt C} is the number of cycles per trace and {\tt
F} is the number of fences per cycle.  The structure of the query
$\Phi$ corresponds to the {\em Head-cycle-free} (HCF) class of {\tt
CNF} theories; hence, the min-model computation falls in the \fp
complexity class
\cite{angiulli2014tractability}.

\noindent
{\bf Determining optimal memory orders of fences.}
The set {\tt min}$\Phi$ gives a sound solution that is optimal only in the 
number of fences.
The function {\tt assignMO} (lines 6,11) assigns the weakest memory
order to the fences in {\tt min}$\Phi$ that is sound.
Let {\tt min-cycles} represent a set of cycles such that every candidate 
fence in the cycles belongs to {\tt min$\Phi$}. 
The {\tt assignMO} function computes memory order for fences 
of {\tt min-cycles} of each trace as follows:
If a cycle $c \in$ {\tt min-cycles} is detected, then its 
fences must form a $\isetSW$ or $\isetDOB$ with an event of $\imm{\tau}$
(since, candidate fences only modify $\setSB$, $\setSW$ and $\setDOB$).
Let $R$ = $ \isetSW \cup \isetDOB$. The scheme to compute fence types is as follows:
\begin{itemize}
\item If a fence $\mathbb{F}$ in a weak cycle $c$ is related to an event $e$ of 
$c$ by $R$ as 
$e{R}\mathbb{F}$, then $\mathbb{F}$ is assigned the memory order \acq;
\item if an event $e$ in $c$ is related to $\mathbb{F}$ as 
${\mathbb{F}{R}}e$ then $\mathbb{F}$ is assigned \rel;
\item if events $e,e'$ of $c$ are related to $\mathbb{F}$ as $e{R}\mathbb{F}{R}e'$ then 
$\mathbb{F}$ is assigned \acqrel.
\item All the fences in a strong cycle are assigned the memory order \sc.
\end{itemize}
Consider a cycle $c$:
$e$$\seqb{\imm{\tau'}}{}{}$$\mathbb{F}_1$$\sw{\imm{\tau'}}{}{}$$\mathbb{F}_2$$\sw{\imm{\tau'}}{}{}$$\mathbb{F}_3$$\seqb{\imm{\tau'}}{}{}$ $e'$$\rf{\imm{\tau'}}{}{}$$e$
representing a violation of $\rf{\imm{\tau'}}{}{};\hb{\imm{\tau'}}{}{}$
irreflexivity (condition \hlref{co-rh}).
According to the scheme discussed above, the fences
$\mathbb{F}_1$, $\mathbb{F}_2$ and $\mathbb{F}_3$ are assigned 
the memory orders \rel, \acqrel and \acq respectively and 
$wt(c)$ = 4 (refer $\mathsection$\ref{sec:preliminaries}).
%
%
			
			
%
Further, {\tt assignMO} iterates over all buggy traces and detects 
the  sound weakest memory order for each fence across 
all traces as follows.
Assume a cycle $c_1$ in $\imm{\tau_1}$ and a 
cycle $c_2$ in $\imm{\tau_2}$. The function computes a union of
the fences of $\tau_1$ and $\tau_2$ while choosing the stronger
memory order for each fence that is present in both the cycles.
In doing so,  both $\tau_1$ and
$\tau_2$ are invalidated. 
Further, when two candidate solutions have the same set of
fences, the function selects the one with the lower weight.

Consider the cycles of buggy traces $\tau_1$ and $\tau_2$
shown in \hlref{candidate-fences}.
Let {\tt min$\Phi$} = $\{\mathbb{F}_1, \mathbb{F}_2, 
\mathbb{F}_3\}$.
The memory orders of the fences for each trace are shown 
with superscripts and the weights of the cycle $\tau_1c_1$,
$\tau_1c_2$ and $\tau_2c_1$ are  
written against the name of the cycles.
%
%
The candidate solutions $\tau_1c_1$ and $\tau_1c_2$ are
combined with $\tau_2c_1$
to form $\tau_{12}c_{11}$ and $\tau_{12}c_{21}$ of
weights $5$ and $4$, respectively. The
solution $\tau_{12}c_{11}$ is of higher weight and is
discarded.
In $\tau_{12}c_{21}$, the optimal memory orders
\rel, \acq and \acqrel are assigned to fences $\mathbb{F}_1$,
$\mathbb{F}_2$ and $\mathbb{F}_3$, respectively.
%
%
%
It is possible that {\tt min-cycles} may contain fences originally in
$P$.  If the process discussed above computes a stronger memory order
for a program fence than its original order in $P$, then the technique
strengthens the memory order of the fence to the computed order.
Note that  this reasoning across traces does not occur in \fasttechnique
as it considers only one trace at a time.

%
%

Determining the optimal memory orders has a complexity in
$\bigo$({\tt BT}.{\tt C}.{\tt F}+{\tt M}$^{\tt BT}$),
where {\tt BT} if the number of buggy traces of $P$, 
{\tt C} and {\tt F} are defined as before, and  
{\tt M} is the number of min-cycles per trace.

In our experimental observation (refer to $\mathsection$\ref{sec:results}), the number
of buggy traces analyzed by \fasttechnique is significantly less than
$|${\tt BT}$|$. Therefore, in practice, the complexity of various
steps of \fasttechnique that are dependent on {\tt BT} reduces
exponentially by a factor of $|${\tt BT}$|$.

%

%
%
%
\noindent {\bf Nonoptimality of \fasttechnique.} 
%
Consider the example \hlref{3-fence}.
It shows cycles in two buggy traces $\tau_1$ and $\tau_2$ of an 
input program.
\begin{figure}[!t]
	\centering
	\begin{minipage}{0.45\textwidth}            
   		\centering
		\setlength{\tabcolsep}{7pt}
		\begin{tabular}{|l|}
			\hline
			$\tau_1c_1$(4): $\mathbb{F}^\acqrel_1$ $\^$ $\mathbb{F}^\acqrel_2$ \\
			$\tau_1c_2$(4): $\mathbb{F}^\rel_1$ $\^$ $\mathbb{F}^\acq_2$ $\^$ $\mathbb{F}^\acqrel_3$ \\
			\multicolumn{1}{|c|}{cycles of $\tau_1$} \\
			\hline
			
			$\tau_2c_1$(3): $\mathbb{F}^\rel_1$ $\^$ $\mathbb{F}^\acq_2$ $\^$ $\mathbb{F}^\acq_3$ \\
			\multicolumn{1}{|c|}{cycle of $\tau_2$} \\
			\hline
			
			$\tau_{12}c_{11}$(5): $\mathbb{F}^\acqrel_1$ $\^$ $\mathbb{F}^\acqrel_2$ $\^$ $\mathbb{F}^\acq_3$ \\
			$\tau_{12}c_{21}$(4): $\mathbb{F}^\rel_1$ $\^$ $\mathbb{F}^\acq_2$ $\^$ $\mathbb{F}^\acqrel_3$ \\
               \hline            
			\multicolumn{1}{c}{\hl{candidate-fences}}\\
		\end{tabular}
	\end{minipage}
	\begin{minipage}{0.45\textwidth}
		\setlength{\tabcolsep}{7pt}
		\begin{tabular}{|l|}
			\hline
			cycles in $\tau_1$ (${\tt C}_{\tau_1}$): \\
			\quad $\{\mathbb{F}_1, \mathbb{F}_2, e_1\}$ and
			$\{\mathbb{F}_1, \mathbb{F}_3, \mathbb{F}_4\}$\\
			$\Phi_{\tau_1}{=}$ ($\mathbb{F}_1 \^ \mathbb{F}_2) {\v} (\mathbb{F}_1 \^ \mathbb{F}_3 \^ \mathbb{F}_4$)\\
			\\
			cycles in $\tau_2$ (${\tt C}_{\tau_2}$):
			$\{\mathbb{F}_3, \mathbb{F}_4\}$\\
			$\Phi_{\tau_2}{=}$ ($\mathbb{F}_3 \^ \mathbb{F}_4$)\\
			\hline
			\multicolumn{1}{c}{\hl{3-fence}} \\
		\end{tabular}
	\end{minipage}
\end{figure}
\ourtechnique provides the formula $\Phi_{\tau_1}$ $\^$ 
$\Phi_{\tau_2}$ to the \sat solver and the optimal solution
obtained is ($\mathbb{F}_1 \^ 
\mathbb{F}_3 \^ \mathbb{F}_4$). 
However, \fasttechnique considers the formula $\Phi_{\tau_1}$ and
$\Phi_{\tau_2}$ in separate iterations and may return a nonoptimal
result ($\mathbb{F}_1 \^ \mathbb{F}_2$) $\^$
($\mathbb{F}_3 \^ \mathbb{F}_4$).

We prove the soundness of \fasttechnique and \ourtechnique with
Theorems~\ref{thm:fastfensying sound} and \ref{thm:fensying sound}
respectively and the optimality of \ourtechnique with
Theorem~\ref{thm:fensying optimal}.

\noindent
{\bf Notations for Theorem proofs.}
For an input program $P$, let $P'$ = $prg(P, \events')$,
represent a transformation of $P$ constructed by adding the 
events $\events'$ to the original events of $P$, where 
$\events'$ is a set of fence events \ie $\forall e \in \events'$
$act(e)$ = fence.
Further, let $bt(P)$ represent the set of buggy traces of $P$. 
Let $\inv{\tau}$ represent the invalidated version of $\tau$ 
with synthesized fences of a candidate solution and let
$\fx{P}$ represent the fixed version of $P$ with no more buggy traces.
Let $\mathbb{C}$ represent the set of relation compositions 
corresponding to the \hlref{coherence conditions} \ie \newline
$\mathbb{C}$ = $\{$ 
($\setHB$), 
($\setRF$;$\setHB$), 
($\setMO$;$\setRF$;$\setHB$), 
($\setMO$;$\setHB$), 
($\setMO$;$\setHB$;$\setRF^{-1}$), 
($\setMO$;$\setRF$;$\setHB$;$\setRF^{-1}$) $\}$.
For a fence $\mathbb{F} \in$ {\tt min}$\Phi$, let $\mathbb{F}^m$
represent the same fence with memory order $m$ assigned by the 
{\tt assignMO} routine (line 14 of Algorithm~\ref{alg:main algo}).

{\lemma {\wkfence is sound: \newline
		{\normalfont
			Given an input program $P$,
			$\forall \tau \in bt(P)$,
			$\exists cond \in \mathbb{C}$ \st $cond$ is reflexive $\iff$
			$weak$-$cycles_\tau$ $\neq \emptyset$.
		}\newline
		There exists a violation of a coherence condition {\em if-and-only-if} 
		\wkfence detects a cycle in the corresponding relation compositions.} 
	\label{lem:weak-sound}}
\begin{proof}
	Case $\implies$: 
$\exists cond \in \mathbb{C}$ \st $cond$ is reflexive $\implies$
$weak$-$cycles_\tau$ $\neq \emptyset$.\newline
\wkfence  checks the validity of coherence conditions
on event relations between program events and synthesized
fences. 
The validity is checked by detecting cycles in relation 
compositions of coherence conditions using {\em Johnson's} algorithm
(refer $\mathsection$\ref{sec:methodology}).
Since, {\em Johnson's} algorithm soundly detects all cycles, \wkfence 
is sound if the event relations $\hb{\imm{\tau}}{}{}$, 
$\rf{\imm{\tau}}{}{}$, $\mo{\imm{\tau}}{}{}$ and $\rfinv{\imm{\tau}}{}{}$
are correctly computed (\ie $\nexists e_1,e_2 \in \events_{\imm{\tau}}$ 
\st a cycle would be formed containing an ordering of
$e_1,e_2$ but the pair is not in the corresponding relation
$\hb{\imm{\tau}}{}{}$ or $\rf{\imm{\tau}}{}{}$ or $\mo{\imm{\tau}}{}{}$ 
or $\rfinv{\imm{\tau}}{}{}$.
(Note that $\fr{\imm{\tau}}{}{}$ relation is not invoked by any 
coherence condition.)

Given a buggy trace $\tau$, we get the relations $\setHB$,
$\setRF$, $\setMO$ and $\setRF^{-1}$ from the buggy trace generator.
We compute the $\hb{\imm{\tau}}{}{}$ relation after introducing 
the synthesized fences in the intermediate trace $\imm{\tau}$, 
hence, the soundness condition can be defined as:\newline 
\ourtechnique soundly detects all weak cycle without recomputing		 
$\setRF$, $\setMO$ and $\setRF^{-1}$ relations for the events of 
$\imm{\tau}$ (\ie $\rf{\imm{\tau}}{}{} = \setRF$, 
$\mo{\imm{\tau}}{}{} = \setMO$ and $\rfinv{\imm{\tau}}{}{} = 
\setRF^{-1}$).

\begin{figure}[h]
	\begin{tabular}{|c|c|c|c|}
		\hline
		\resizebox{0.17\textwidth}{!}{\tikzset{every picture/.style={line width=0.75pt}} 
\begin{tikzpicture}[x=1em,y=1em,yscale=-1,xscale=-1]
\tikzstyle{every node}=[font=\normalfont]
\node (w1) {$ w_1 $};
\node (w2) [below=20pt of w1] {$ w_2 $};
\node (coww) [below=-3pt of w2] {\hl{coWW}};

\draw [->,>=stealth,color=Cerulean] ($ (w1.south east)+(.3,-5pt) $) to[out=135,in=-125] node[midway,right=-2pt,font=\scriptsize] {\textcolor{black}{\lhb}} ($ (w2.north east)+(0.2,3pt) $);
\draw [->,>=stealth,color=RedOrange] ($ (w2.north west)+(-0.2,3pt) $) to[out=-45,in=45] node[left=-2pt,pos=.25,font=\scriptsize] {\textcolor{black}{\lmo}}($ (w1.south west)+(-0.3,-5pt) $);

\end{tikzpicture}} &
		\resizebox{0.22\textwidth}{!}{\tikzset{every picture/.style={line width=0.75pt}} 
\begin{tikzpicture}[x=1em,y=1em,yscale=-1,xscale=-1]
\tikzstyle{every node}=[font=\normalfont]

\node (w1) [inner sep=2pt] {$ w_1 $};
\node (r1) [right=35pt of w1,inner sep=2pt] {$ r_1 $};
\node (w2) [below=20pt of w1,inner sep=2pt] {$ w_2 $};
\node (cowr) [below right=-4pt and 5pt of w2] {\hl{coWR}};

\draw [->,>=stealth,color=Cerulean] (w1) -- node[midway,above=-2pt,font=\scriptsize,color=black] { $\lhb$ } (r1);
\draw [->,>=stealth,color=PineGreen] (r1) -- node[midway,below=-1pt,font=\scriptsize,color=black]{$\lrf^{-1}$} (w2);
\draw [->,>=stealth,color=RedOrange] (w2) -- node[midway,left=-2pt,font=\scriptsize,color=black] { $\lmo$ } (w1);

\end{tikzpicture}} &
		\resizebox{0.22\textwidth}{!}{\tikzset{every picture/.style={line width=0.75pt}} 
\begin{tikzpicture}[x=1em,y=1em,yscale=-1,xscale=-1]
\tikzstyle{every node}=[font=\normalfont]

\node (w1) [inner sep=2pt] {$ w_1 $};
\node (r1) [right=35pt of w1,inner sep=2pt] {$ r_1 $};
\node (w2) [below=20pt of w1,inner sep=2pt] {$ w_2 $};
\node (corw) [below right=-4pt and 5pt of w2] {\hl{coRW}};

\draw [->,>=stealth,color=Cerulean] (r1) -- node[midway,below=-1pt,font=\scriptsize,color=black] { $\lhb$ } (w2);
\draw [->,>=stealth,color=PineGreen] (w1) -- node[midway,above=-2pt,font=\scriptsize,color=black]{\lrf} (r1);
\draw [->,>=stealth,color=RedOrange] (w2) -- node[midway,left=-2pt,font=\scriptsize,color=black] { $\lmo$ } (w1);

\end{tikzpicture}} &
		\resizebox{0.24\textwidth}{!}{\tikzset{every picture/.style={line width=0.75pt}} 
\begin{tikzpicture}[x=1em,y=1em,yscale=-1,xscale=-1]
\tikzstyle{every node}=[font=\normalfont]

\node (w1) [inner sep=2pt] {$ w_1 $};
\node (r1) [right=35pt of w1,inner sep=2pt] {$ r_1 $};
\node (w2) [below=20pt of w1,inner sep=2pt] {$ w_2 $};
\node (r2) [below=20pt of r1,inner sep=2pt] {$ r_2 $};
\node (corr) [below right=-2pt and 5pt of w2] {\hl{coRR}};

\draw [->,>=stealth,color=Cerulean] (r1) -- node[midway,right=-2pt,font=\scriptsize,color=black] { $\lhb$ } (r2);
\draw [->,>=stealth,color=PineGreen] (w1) -- node[midway,above=-2pt,font=\scriptsize,color=black]{\lrf} (r1);
\draw [->,>=stealth,color=PineGreen] (r2) -- node[midway,above=-2pt,font=\scriptsize,color=black]{$\lrf^{-1}$} (w2);
\draw [->,>=stealth,color=RedOrange] (w2) -- node[midway,left=-2pt,font=\scriptsize,color=black] { $\lmo$ } (w1);

\end{tikzpicture}} \\
		\hline
	\end{tabular}
	\label{fig:como}
\end{figure}

\begin{itemize}[label=setmm,align=left,leftmargin=*]
	\item [$\setRF$] The relation is formed from write (or rmw) events 
	to read (or rmw) events, since fences cannot be both, the $\setRF$
	relations remains unchanged \ie $\rf{\imm{\tau}}{}{}$ =
	$\setRF$.
	
	\item [$\setRF^{-1}$] The relation remains unchanged as 
	$\setRF$ remains unchanged \ie $\rfinv{\imm{\tau}}{}{}$ 
	= $\setRF^{-1}$.
	
	\item [$\setMO$] Assume $\exists w, w' \in \events_\tau$ \st 
	as a consequence of synthesizing fences in the buggy trace 
	$\tau$ to form $\imm{\tau}$, $w$ is {\em modification-ordered
		before} $w'$. However, $(w,w') \nin \mo{\imm{\tau}}{}{}$
	since we consider $\mo{\imm{\tau}}{}{}$ = $\setMO$.
	
	We show by case analysis on the coherence conditions
	involving $\mo{\imm{\tau}}{}{}$ that \ourtechnique does
	not miss a weak cycle by not expanding the modification-order
	after synthesizing fences.
	
	Consider the following four cases of coherence involving 
	$\mo{\imm{\tau}}{}{}$ (borrowed from 
	\cite{batty2011mathematizing}):
	
	\begin{itemize}[label=CoWW,align=left,leftmargin=*]
		\item [CoWW:] 
		$\forall$ \cc traces $\tau^{c11}$,
		$\nexists w_1,w_2 \in \writes_{\tau^{c11}}$ \st
		$\hb{\tau^{c11}}{w_1}{w_2}$ and 
		$\mo{\tau^{c11}}{w_2}{w_1}$.

		Now, let $\exists w_1, w_2 \in \writes$
		\st $\hb{\imm{\tau}}{w_1}{w_2}$. 
		
		If $\mo{\tau}{w_1}{w_2}$ then there does not
		exist a violation.
		
		However, if $\mo{\tau}{w_2}{w_1}$ then we will 
		detect the violation as a cycle in 
		$\mo{\imm{\tau}}{}{}$;$\hb{\imm{\tau}}{}{}$
		(depicted diagrammatically in \hlref{coWW}).
		
		\item [CoWR:] 
		$\forall$ \cc traces $\tau^{c11}$,
		$\nexists w_1,w_2 \in \writes_{\tau^{c11}}$,
		$r_1 \in \reads_{\tau^{c11}}$ \st
		$\hb{\tau^{c11}}{w_1}{r_1}$,
		$\mo{\tau^{c11}}{w_2}{w_1}$ and
		$\rf{\tau^{c11}}{w_2}{r_1}$.
		
		Now, let $\exists r_1 \in \reads$, 
		$\exists w_1,w_2 \in \writes$ 
		\st $\hb{\imm{\tau}}{w_1}{r_1}$ and 
		$\rf{\imm{\tau}}{w_2}{r_1}$.
		
		If $\mo{\tau}{w_1}{w_2}$ then there does not
		exist a violation.
		
		However, if $\mo{\tau}{w_2}{w_1}$ then we will 
		detect the violation as a cycle in 
		$\mo{\imm{\tau}}{}{}$;$\hb{\imm{\tau}}{}{}$;$\rfinv{\imm{\tau}}{}{}$
		(depicted diagrammatically in \hlref{coWR}).
		
		\item [CoRW:] 
		$\forall$ \cc traces $\tau^{c11}$,
		$\nexists w_1,w_2 \in \writes_{\tau^{c11}}$,
		$r_1 \in \reads_{\tau^{c11}}$ \st
		$\hb{\tau^{c11}}{r_1}{w_2}$,
		$\mo{\tau^{c11}}{w_2}{w_1}$ and
		$\rf{\tau^{c11}}{w_1}{r_1}$.
		
		Now, let $\exists r_1 \in \reads$, 
		$\exists w_1,w_2 \in \writes$ 
		\st $\rf{\imm{\tau}}{w_1}{r_1}$ and 
		$\hb{\imm{\tau}}{r_1}{w_2}$.
		
		If $\mo{\tau}{w_1}{w_2}$ then there does not
		exist a violation.
		
		However, if $\mo{\tau}{w_2}{w_1}$ then we will 
		detect the violation as a cycle in 
		$\mo{\imm{\tau}}{}{}$;$\rf{\imm{\tau}}{}{}$;$\hb{\imm{\tau}}{}{}$
		(depicted diagrammatically in \hlref{coRW}).
		
		\item [CoRR:] 
		$\forall$ \cc traces $\tau^{c11}$,
		$\nexists w_1,w_2 \in \writes_{\tau^{c11}}$,
		$r_1,r_2 \in \reads_{\tau^{c11}}$ \st
		$\hb{\tau^{c11}}{r_1}{r_2}$,
		$\mo{\tau^{c11}}{w_2}{w_1}$,
		$\rf{\tau^{c11}}{w_1}{r_1}$ and
		$\rf{\tau^{c11}}{w_2}{r_2}$.
		
		Now, let $\exists r_1,r_2 \in \reads$,
		$\exists w_1, w_2 \in \writes$
		\st $\rf{\imm{\tau}}{w_1}{r_1}$, 
		$\rf{\imm{\tau}}{w_2}{r_2}$ and
		$\hb{\imm{\tau}}{r_1}{r_2}$. 
		
		If $\mo{\tau}{w_1}{w_2}$ then there does not
		exist a violation.
		
		However, if $\mo{\tau}{w_2}{w_1}$ then we will 
		detect the violation as a cycle in 
		$\mo{\imm{\tau}}{}{}$; $\rf{\imm{\tau}}{}{}$; $\hb{\imm{\tau}}{}{}$; $\rfinv{\imm{\tau}}{}{}$
		(depicted diagrammatically in \hlref{coRR}).
	\end{itemize}
\end{itemize}

Thus, \ourtechnique does not miss a cycle in any coherence rule
that is violated.\newline

Case $\impliedby$:	
$\exists cond \in \mathbb{C}$ \st $cond$ is reflexive 
$\impliedby$
$weak$-$cycles_\tau$ $\neq \emptyset$.\newline
\ourtechnique expands $\setHB$ to $\hb{\imm{\tau}}{}{}$, while
$\rf{\imm{\tau}}{}{} = \setRF$, $\mo{\imm{\tau}}{}{} = \setMO$ and 
$\rfinv{\imm{\tau}}{}{} = \setRF^{-1}$.
The computation of $\hb{\imm{\tau}}{}{}$ is borrowed from
\cc standard \cite{C11,batty2011mathematizing} and thus, is sound.
\newline
Further, {\em Johnson's} algorithm (used for detecting cycles in 
the computed event relations) is sound. Thus, if the algorithm
returns a cycle $eCe$ where $e \in \events_{\imm{\tau}}$ and
$C \in \mathbb{C}$ then the condition $C$ is reflexive.

Hence, if \ourtechnique detects a weak cycle then a corresponding
coherence condition is violated.
\end{proof}

{\lemma {$\setSO^+ \subseteq \setTO$\newline 
		For any valid \cc trace $\tau$, each pair of events related by
		$\setSO$ are ordered by $\setTO$.\newline
		In other words,
		$\setSO$ does not order events if the ordering violates \hlref{to-sc}.}
	\label{lem:so subset to}}
\begin{proof}
	By definition of $\setSO$, $\so{\tau}{e^\sc_1}{e^\sc_2}$ where\newline
$(e^\sc_1, e^\sc_2) \in$ $\setHB$ $\union$ 
$\setMO$ $\union$ $\setRF$ $\union$ $\setFR$, and
\begin{itemize}[label=soee,align=left,leftmargin=*]
	\item [soee:] $e^\sc_1, e^\sc_2 \in \ordevents{\sc}_\tau$
	$\implies$ $\to{\tau}{e^\sc_1}{e^\sc_2}$
	(since $\to{\tau}{e^\sc_2}{e^\sc_1}$ violates 
	\hlref{coto} or \hlref{rfto1}).
	
	\item [soef:] $e^\sc_1 \in \ordevents{\sc}_\tau$,
	$e^\sc_2 \in \ordfences{\sc}_\tau$
	$\implies$ $\to{\tau}{e^\sc_1}{e^\sc_2}$
	(since $\to{\tau}{e^\sc_2}{e^\sc_1}$ violates 
	\hlref{coto} or \hlref{rfto1}).
	
	\item [sofe:] $e^\sc_2 \in \ordevents{\sc}_\tau$,
	$e^\sc_1 \in \ordfences{\sc}_\tau$
	$\implies$ $\to{\tau}{e^\sc_1}{e^\sc_2}$
	(since $\to{\tau}{e^\sc_2}{e^\sc_1}$ violates 
	\hlref{coto} or \hlref{rfto1} or \hlref{rfto2}).
	
	\item [soff:] $e^\sc_1, e^\sc_2 \in \ordfences{\sc}_\tau$
	$\implies$ $\to{\tau}{e^\sc_1}{e^\sc_2}$
	(since $\to{\tau}{e^\sc_2}{e^\sc_1}$ violates 
	\hlref{coto} or \hlref{frfto1}).
\end{itemize}
Since, $\setTO$ is total, thus, $\setSO^+ \subseteq \setTO$.
\end{proof}

{\lemma {\stfence is sound:\newline 
		$\neg$($total(\ordevents{\sc}, \to{\imm{\tau}}{}{})$ $\^$ 
		$order(\ordevents{\sc}, \to{\imm{\tau}}{}{})$) $\iff$
		$strong$-$cycles_\tau$ $\neq \emptyset$. \newline
		There does not exist a total order on the \sc ordered events 
		of an intermediate trace $\imm{\tau}$ {\em if-and-only-if}
		there exists a cycle in $\so{\imm{\tau}}{}{}$.} 
	\label{lem:strong-sound}}
\begin{proof}
		Case $\implies$: 
$\neg$($total(\ordevents{\sc}, \to{\imm{\tau}}{}{})$ $\^$ 
$order(\ordevents{\sc}, \to{\imm{\tau}}{}{})$) $\implies$
$strong$-$cycles_\tau$ $\neq \emptyset$.\newline
Consider $e_1, e_2 \in \ordevents{\sc}_\tau$ \st 
both $\to{\tau}{e_1}{e_2}$ and 
$\to{\tau}{e_2}{e_1}$ do not violate the conditions \hlref{coto}, 
\hlref{rfto1}, \hlref{rfto2} and \hlref{frfto}.
To form the total order we can assume either one of the two orders
\cite{C11}. Assume $\to{\tau}{e_1}{e_2}$.

Further, consider a total order cannot be formed on \sc events of
$\imm{\tau}$ \st
$e$ $\to{\imm{\tau}}{}{}$ $...$ $\to{\imm{\tau}}{}{}$ $e_1$ 
$\to{\imm{\tau}}{}{}$ $e_2$ $\to{\imm{\tau}}{}{}$ $...$ 
$\to{\imm{\tau}}{}{}$ $e$
then we simply flip $\to{\imm{\tau}}{e_1}{e_2}$ to 
$\to{\imm{\tau}}{e_2}{e_1}$ and eliminate the cycle.

Further, if a cycle in $\to{\imm{\tau}}{}{}$ includes 
$\to{\imm{\tau}}{e_1}{e_2}$ and another cycle includes 
$\to{\imm{\tau}}{e_2}{e_1}$ then there exists a cycle 
$\to{\imm{\tau}}{e_1}{\to{\imm{\tau}}{...}{\to{\imm{\tau}}{e_2}{\to{\imm{\tau}}{...}{e_1}}}}$
(by by definition of $\to{\imm{\tau}}{}{}$, as shown in the figure below).
Thus, pairs of \sc ordered events that don't have a fixed $\setTO$ order
cannot contribute to a strong cycle. \hfill$\inf$(i).

\begin{figure}[h]
	\resizebox{0.2\textwidth}{!}{\tikzset{every picture/.style={line width=0.75pt}} 
\begin{tikzpicture}[x=1em,y=1em,yscale=-1,xscale=-1]
\tikzstyle{every node}=[font=\normalfont]
\node (e1) {$ e_1 $};
\node (e2) [below=20pt of w1] {$ e_2 $};

\draw [->,>=stealth,dashed,color=Mahogany] ($ (e1.south)+(-0.2,0) $) -- ($ (e2.north)+(-0.2,0) $);
\draw [->,>=stealth,dashed,color=Mahogany] ($ (e2.north)+(0.2,0) $) -- ($ (e1.south)+(0.2,0) $);

\draw [->,>=stealth,color=Mahogany] ($ (e1.north east)+(.3,3pt) $) to[out=-155,in=155] ($ (e2.south east)+(0.2,-3pt) $);
\draw [->,>=stealth,color=Mahogany] ($ (e2.south west)+(-0.2,-3pt) $) to[out=25,in=-25] ($ (e1.north west)+(-0.3,4pt) $);

\end{tikzpicture}}
\end{figure}

Now, if there does not exist a total order on the \sc ordered events of 
$\imm{\tau}$ then 

$\neg$($total(\ordevents{\sc}, \to{\imm{\tau}}{}{})$
$\^$ $order(\ordevents{\sc}, \to{\imm{\tau}}{}{})$), \ie

$\neg$($total(\ordevents{\sc}, \to{\imm{\tau}}{}{})$) $\v$
$\exists e \in \ordevents{\sc}_{\imm{\tau}}$ \st $\to{\imm{\tau}}{e}{e}$
$\v$ $\neg(\reln{\textbf{\textcolor{Brown}{to}}^+}{\imm{\tau}}{}{}
\subseteq \to{\imm{\tau}}{}{})$ 

\hfill(by definition of 
$order(\ordevents{\sc}, \to{\imm{\tau}}{}{})$).

By definition of $\to{\imm{\tau}}{}{}$, 
$\neg$$total(\ordevents{\sc}, \to{\imm{\tau}}{}{})$ and 
$\neg(\reln{\textbf{\textcolor{Brown}{to}}^+}{\imm{\tau}}{}{}
\subseteq \to{\imm{\tau}}{}{})$ are not feasible.

\noindent
Thus, $\exists e \in \ordevents{\sc}_{\imm{\tau}}$ \st $\to{\imm{\tau}}{e}{e}$
\newline
$\implies$
\sc events of $\imm{\tau}$ violate \hlref{coto}, \hlref{rfto1}, 
\hlref{rfto2} or \hlref{frfto}. 

\begin{itemize}[label=aafrfto,align=left,leftmargin=*]
	\item [{[coto]}]
	Let $\exists e^\sc \in \ordevents{\sc}_{\imm{\tau}}$ \st
	$(e,e) \in$ ($\isetHB$ $\union$ $\isetMO$).
	Thus, \hlref{coto} is violated by $e^\sc$.
	
	As we know that, $\onsc{\isetMO}$ $\union$ $\onsc{\isetHB}$
	$\subseteq$ $\isetSO^+$ thus we have a cycle 
	$\so{\imm{\tau}}{e^\sc}{\so{\imm{\tau}}{...}{e^\sc}}$.
	
	\item [{[rfto1]}]
	Let $\exists w_1^\sc, w_2^\sc \in \ordwrites{\sc}_{\imm{\tau}}$,
	$r_1^\sc \in \ordreads{\sc}_{\imm{\tau}}$ \st
	$\to{\imm{\tau}}{w_1^\sc}{\to{\imm{\tau}}{w_2^\sc}{r_1^\sc}}$ and
	$\rf{\imm{\tau}}{w_1^\sc}{r_1^\sc}$.\newline
	Thus, \hlref{rfto1} is violated by $w_1^\sc$, $w_2^\sc$ and $r_1^\sc$.
	
	Since, $\tau$ is a valid trace, $\neg\to{\tau}{w^\sc_1}{w^\sc_2}$ $\v$
	$\neg\to{\tau}{w^\sc_2}{r_1^\sc}$.
	
	Further, since inserting fences only modifies the $\setHB$ relation,
	if $\neg\to{\tau}{w^\sc_1}{w^\sc_2}$ then $\hb{\imm{\tau}}{w^\sc_1}{w^\sc_2}$
	(because $\to{\imm{\tau}}{w^\sc_1}{w^\sc_2}$). 
	Similarly, 
	if $\neg\to{\tau}{w^\sc_2}{r^\sc_1}$ then $\hb{\imm{\tau}}{w^\sc_2}{r^\sc_1}$.
	
	Also, $\to{\imm{\tau}}{w^\sc_1}{w^\sc_2}$ $\implies$ $\mo{\imm{\tau}}{w^\sc_1}{w^\sc_2}$
	(assuming \hlref{co-mh} is not violated) 
	$\implies$ $\fr{\imm{\tau}}{r^\sc_1}{w^\sc_2}$.
	
	As we know that, $\onsc{\isetFR}$ $\union$ $\onsc{\isetHB}$
	$\subseteq$ $\isetSOtr$ thus we have a cycle 
	$\so{\imm{\tau}}{r^\sc_1}{\so{\imm{\tau}}{w^\sc_2}{r^\sc_1}}$.
	
	\item [{[rfto2]}]
	Let $\exists w_1, w_2^\sc \in \writes_{\imm{\tau}}$, 
	$r_1^\sc \in \ordreads{\sc}_{\imm{\tau}}$ \st $ord(w_2^\sc)$ is \sc,
	$\rf{\imm{\tau}}{w_1}{r_1^\sc}$, $\hb{\imm{\tau}}{w_1}{w_2^\sc}$ and
	$\to{\imm{\tau}}{w_2^\sc}{r_1^\sc}$.\newline
	Thus, \hlref{rfto2} is violated by $w_1$, $w_2^\sc$ and $r_1^\sc$.
	
	Since, $\tau$ is a valid trace, $\neg\hb{\tau}{w_1}{w^\sc_2}$ $\v$
	$\neg\to{\tau}{w^\sc_2}{r_1^\sc}$.
	
	Further, since inserting fences only modifies the $\setHB$ relation,
	if $\neg\to{\tau}{w^\sc_2}{r^\sc_1}$ then $\hb{\imm{\tau}}{w^\sc_2}{r^\sc_1}$
	(because $\to{\imm{\tau}}{w^\sc_2}{r^\sc_1}$). 
	
	Also, $\hb{\imm{\tau}}{w_1}{w^\sc_2}$ $\implies$ $\mo{\imm{\tau}}{w_1}{w^\sc_2}$
	(assuming \hlref{co-mh} is not violated) 
	$\implies$ $\fr{\imm{\tau}}{r^\sc_1}{w^\sc_2}$.
	
	As we know that, $\onsc{\isetFR}$ $\union$ $\onsc{\isetHB}$
	$\subseteq$ $\isetSOtr$ thus we have a cycle 
	$\so{\imm{\tau}}{r^\sc_1}{\so{\imm{\tau}}{w^\sc_2}{r^\sc_1}}$.
	
	\item [{[frfto]}]
	Let $\exists w_1^\sc, w_2^\sc \in \ordwrites{\sc}_{\imm{\tau}}$,
	$r_1 \in \reads_{\imm{\tau}}$, 
	$\mathbb{F}^\sc \in \ordfences{\sc}_{\imm{\tau}}$ \st
	$\to{\imm{\tau}}{w_1^\sc}{\to{\imm{\tau}}{w_2^\sc}{\mathbb{F}^\sc}}$,
	$\seqb{\imm{\tau}}{\mathbb{F}^\sc}{r_1}$ and
	$\rf{\imm{\tau}}{w_1^\sc}{r_1}$.\newline
	Thus, \hlref{frfto} is violated by $w_1^\sc$, $w_2^\sc$, $r_1$ and
	$\mathbb{F}^\sc$.
	
	Since, $\tau$ is a valid trace, $\neg\to{\tau}{w^\sc_1}{w^\sc_2}$ $\v$
	$\neg\to{\tau}{w^\sc_2}{\mathbb{F}^\sc}$.
	
	Further, since inserting fences only modifies the $\setHB$ relation,
	if $\neg\to{\tau}{w^\sc_1}{w^\sc_2}$ then $\hb{\imm{\tau}}{w^\sc_1}{w^\sc_2}$
	(because $\to{\imm{\tau}}{w^\sc_1}{w^\sc_2}$). 
	Similarly, if $\neg\to{\tau}{w^\sc_2}{\mathbb{F}^\sc}$ then 
	$\hb{\imm{\tau}}{w^\sc_2}{\mathbb{F}^\sc}$.
	
	Also, $\to{\imm{\tau}}{w^\sc_1}{w^\sc_2}$ $\implies$ $\mo{\imm{\tau}}{w^\sc_1}{w^\sc_2}$
	(assuming \hlref{co-mh} is not violated) 
	$\implies$ $\fr{\imm{\tau}}{r_1}{w^\sc_2}$
	$\implies$ $\so{\imm{\tau}}{\mathbb{F}^\sc}{w^\sc_2}$ (using \hlref{sofe}).
	
	As we know that, $\onsc{\isetHB}$
	$\subseteq$ $\isetSO$ thus we have a cycle 
	$\so{\imm{\tau}}{r^\sc_1}{\so{\imm{\tau}}{w^\sc_2}{r^\sc_1}}$.
\end{itemize}

Case $\impliedby$: 
$\neg$($total(\ordevents{\sc}, \to{\imm{\tau}}{}{})$ $\^$ 
$order(\ordevents{\sc}, \to{\imm{\tau}}{}{})$) $\impliedby$
$strong$-$cycles_\tau$ $\neq \emptyset$.\newline
$strong$-$cycles_\tau$ $\neq \emptyset$ $\implies$ $\exists e 
\in \events_{\imm{\tau}}$ \st 
$\SOtr{\imm{\tau}}{e}{e}$
$\implies$ $\to{\imm{\tau}}{e}{e}$ (using Lemma~\ref{lem:so subset to}).
Thus, 
$\neg$($order(\ordevents{\sc}, \to{\imm{\tau}}{}{})$) $\implies$
$\neg$($total(\ordevents{\sc}, \to{\imm{\tau}}{}{})$ $\^$ 
$order(\ordevents{\sc}, \to{\imm{\tau}}{}{})$).

%
%
\end{proof}

{\lemma {{\tt AssignMO} is sound: \newline 
		$\forall$ cycles $c$ = $eR_1...e_1R_2\mathbb{F}R_3e_2...R_4e$
		$\in$ {\tt min-cycles} $($where
		$R_i$ $\in$ $\{\isetHB$, $\isetMO$, $\isetRF$, $\isetRF^{-1}$, 
		$\isetSO\})$,
		$e_1R_2\mathbb{F}^mR_3e_2$.\newline
		If a min-cycle $c$ is formed due to event relations introduced 
		by a fence $\mathbb{F}$ then after assigning a 
		memory order $m$ for $\mathbb{F}$ using {\tt assignMO} the 
		event relations still hold; \ie \newline
		\ourtechnique does not assign a memory
		order to a fence that is too weak to stop the buggy trace.}
	\label{lem:mo-sound}}
\begin{proof}
	By definition of \hlref{coherence conditions} if there exists
a weak cycle in the intermediate trace $\imm{\tau}$ then 
$\exists$ $\sw{\imm{\tau}}{e}{e'}$ $\v$ 
$\dob{\imm{\tau}}{e}{e'}$ 

\st $\neg\sw{\tau}{e}{e'}$ $\^$ $\neg\dob{\tau}{e}{e'}$ 
(since, the buggy trace $\tau$ was returned by the 
buggy trace generator).\newline
By definitions of $\sw{\imm{\tau}}{e}{e'}$ and 
$\dob{\imm{\tau}}{e}{e'}$
if $e$ is a fence then its memory order must be \rel or 
stronger, if $e'$ is a fence then its memory order must be 
\acq or stronger. \newline
Since, {\tt assignMO} assigns \rel for $e$ and \acq for $e'$,
thus, the locally assigned memory orders are sufficiently 
strong.\hfill{\it inf}(i)\newline

\noindent
If there exists a fence, $\mathbb{F}$,  that was locally 
assigned a memory order $m$ and after coalescing with other 
buggy trace the final memory order of $\mathbb{F}$ is $m'$
then either $m' = m$ or $m'$ is stronger than $m$
(by construction of coalesced candidate solutions).

\noindent
Since we know that $m$ was sufficiently strong 
(using {\it inf}(i)) 
then the final memory order $m'$ is also sufficiently strong.	
\end{proof}

{\lemma {\ourtechnique is sound for 1 trace.\newline
		Given an input program $P$ \st $bt(P) = \{\tau\}$. 
		$\exists \events'$ \st $bt(prg(P, \events')) = \emptyset$
		$\implies$ \ourtechnique can construct $\inv{\tau}$.
	}
	\label{lem:sound for 1}
}
{\theorem  {\fasttechnique is sound}
	Given an input program $P$, $\exists \tau \in bt(P)$. 
	\st
	$\exists \events'$ \st $bt(prg(P, \events')) = \emptyset$
	$\implies$ \fasttechnique can construct $\inv{\tau}$.
	\label{thm:fastfensying sound}
}
\begin{proof}
	Firstly,
using Lemma~\ref{lem:weak-sound} and Lemma~\ref{lem:strong-sound}
we can state that 
(i) a violation in \hlref{coherence conditions} or \sc total order 
is not missed by \ourtechnique/\fasttechnique, and
(ii) a violation detected by \ourtechnique/\fasttechnique is indeed 
a true violation of either one of the \hlref{coherence conditions} 
or \hlref{tosc}. \newline
Secondly,
the fences introduced for at least 1 of the violations
exist in the final solution (by construction of \sat query).
\ie

$\exists c \in$ $weak$-$cycles_\tau$ $\union$ $strong$-$cycles_\tau$
\st $\forall \mathbb{F}$ (fences included in $c$) $\in$ 
$\fences_{\imm{\tau}} {\setminus}\fences_\tau$ $\mathbb{F}$
$\in$ {\tt min}$\Phi$.\newline
Thirdly,
The memory order assigned to the fences in {\tt min}$\Phi$ is
sufficiently strong to stop the buggy trace 
(Lemma~\ref{lem:mo-sound}($\inf$(i))). \newline
Hence, \ourtechnique is sound for 1 trace, and
\fasttechnique is sound.\newline

\end{proof}

{\theorem {\ourtechnique is sound. \newline
		Given an input program $P$ \st $bt(P) \neq \emptyset$. 
		$\exists \events'$ \st $bt(prg(P, \events')) = \emptyset$
		$\implies$ $\forall \tau \in bt(P)$ we can construct $\inv{\tau}$.
		If $P$ can be fixed by synthesizing or strengthening \cc fences 
		then \ourtechnique fixes $P$.}
	\label{thm:fensying sound}
}
\begin{proof}
	Let {\tt BT} = $bt(P)$.
Consider induction on $|${\tt BT}$|$. 

\noindent
{\sl Base Case:} Consider $|${\tt BT}$|$ = 1. 
Let {\tt BT} = $\{\tau\}$\newline
Using Lemma~\ref{lem:sound for 1},
\ourtechnique is sound for 1 trace.\newline

\noindent 
{\sl Induction Hypothesis:}
Assume that \ourtechnique is sound for $|${\tt BT}$|$ = N.
\newline

\noindent
{\sl Induction Step:}
Consider $|${\tt BT}$|$ = N+1.

Since, we take a conjunction on the \sat formulas from various
traces thus at least 1 cycle from each trace exists in the 
{\tt min-model} (by construction of \sat query).\newline
Further, we know from Lemma~\ref{lem:mo-sound} that \ourtechnique
assigns memory orders that can stop all the corresponding traces.
\newline
Thus, \ourtechnique is sound for N+1 buggy traces.
\end{proof}

{\lemma {{\tt min-model} returns the optimal number of fences.}
	\label{lem:opt-num}}
Let $\fx{\fences}$ represent the set of synthesized fences of 
$\fx{P}$ (\ie $\fx{\fences}$ = $\fences{\setminus}\fences_{\fx{P}}$,
where $\fences_{\fx{P}}$ represents the set of fences of $\fx{P}$) 
then $\nexists \events^o$ \st $|\events^o|$ $<$ 
$|\fx{\fences}|$ and $bt(prg(P,\events^o))$ = $\emptyset$.
\begin{proof}
	Let $\mathcal{F}$ represent the set of fences returned by
{\tt min-model} and let $\mathcal{F}^o$ represent the optimal set
of fences. Assume $|\mathcal{F}^o| < |\mathcal{F}|$.

The min-model of the nonoptimal solution is computed using a
SAT solver (\z) and the computation is assumed to be correct. 
As the consequence,
$|\mathcal{F}^o| < |\mathcal{F}|$ $\implies$ the optimal result
was not a part of the SAT query formula.

Using Lemma~\ref{lem:weak-sound} and 
Lemma~\ref{lem:strong-sound} we know that \ourtechnique does 
not miss any weak or strong cycle
$\implies$ every set of fences that forms a correct solution,
including the optimal solution, is contained in the SAT query
formula.

Thus, by contradiction, $|\mathcal{F}| = |\mathcal{F}|^o$ 
\ie  {\tt min-model} returns the optimal number of fences.
\end{proof}

{\theorem {\ourtechnique synthesizes the optimal number of fences
		with the optimal memory orders.}
	Let $\fx{\fences}$ represent the set of synthesized fences of 
	$\fx{P}$ (\ie $\fx{\fences}$ = $\fences_{\fx{P}}{\setminus}\fences$,
	where $\fences_{\fx{P}}$ represents the set of fences of $\fx{P}$) 
	then $\nexists \events^o$ \st $|\events^o|$ $<$ $|\fx{\fences}|$ or
	($\exists e^o \in \events^o, e \in \fx{\fences}$ \st
	$thr(e^o) = thr(e)$, $idx(e^o) = idx(e)$, $act(e^o) = act(e)$,
	$obj(e^o) = obj(e)$ and $loc(e^o) = loc(e)$ but
	$ord(e^o) \mole ord(e)$) and $bt(prg(P,\events^o))$ = $\emptyset$.
	\label{thm:fensying optimal}
}
\begin{proof}
	We know that {\tt AssignMO} iterates over cycles in {\tt 
	min-cycles} and takes union over fences of cycles from 
{\tt min-cycles}. As {\tt min$\Phi$} consists of the optimal
number of fences (Lemma~\ref{lem:opt-num}) then union over
cycles of {\tt min-cycles} has the same set of fences as
{\tt min$\Phi$}.\newline
Thus, \ourtechnique is optimal in the number of fences.
\newline

Let {\tt BT} represent the set of buggy traces.
Consider induction on $|${\tt BT}$|$. \newline

\noindent
{\sl Base Case:} Consider $|${\tt BT}$|$ = 1.
By definition of {\tt AssignMO} each fence is locally assigned
the weakest memory order that is sound (Lemma~\ref{lem:mo-sound}).

Thus, \ourtechnique is optimal in the memory order of fences
for 1 buggy trace.
\newline

%
%
%
%
%

\noindent
{\sl Induction Hypothesis:} Assume, \ourtechnique is optimal in the 
memory order of fences when $|${\tt BT}$|$ = N.
\newline

\noindent
{\sl Induction Step:} 
Consider $|${\tt BT}$|$ = N+1.\newline
Let $s_1, ..., s_M$ represent the $M$ coalesced solutions for 
buggy traces $\tau_1, ..., \tau_N$ and $\tau_{N+1}c_i$ for 
$i \in \{1, ..., q \}$ represent the $q$ cycles of $(N+1)^{th}$
trace.

Every coalesced solution $\tau_{N+1}s_j$ has the same number of 
fences = fences of {\tt min$\Phi$} because {\tt min$\Phi$} 
returns the minimum number of fences required to stop 
$\tau_1, ..., \tau_{N+1}$.

If there exists a fence $\mathbb{F}$ with memory order $m$ in a 
cycle $\tau_{N+1}c_i$ but the final solution of \ourtechnique
assigns memory order $m'$ to $\mathbb{F}$ \st $m'$ is stronger
than $m$

then, $\exists s_j$ where memory order of $\mathbb{F}$ is
$m'$ (by construction of coalesced solutions),

further, $\nexists s_k$ where memory order of $\mathbb{F}$
is $m$ \st $wt(\tau_{N+1}c_i$-$s_k) < wt(\tau_{N+1}c_i$-$s_j)$
(where $wt(x$-$y)$ represents the weight of the solution formed
by coalescing cycle $x$ with candidate solution $y$).

\noindent
Thus, \ourtechnique is optimal in the memory order of fences
for N+1 buggy traces.
\end{proof}
\section{Implementation and Results} \label{sec:results}
\noindent
{\bf Implementation details.}
The techniques are implemented in {\tt Python}.
\wkfence and \stfence use {\em Johnson's} cycle detection algorithm 
in the {\em networkx} library.
We use \z theorem prover to find the {\em min-model} of \sat queries.
As a \btg, we use \cds~\cite{cds}, an open-source model checker,
for the following reasons;
\begin{enumerate}[noitemsep,topsep=0pt]
	\item \cds supports the \cc semantics. Most other techniques are
	designed for a variant \cite{genmc-PLDI19} or subset
	\cite{abdulla2019verification,tracer2018,singh2021dynamic} of \cc.
	
	\item \cds returns buggy traces 
	along with the corresponding $\setHB$, $\setRF$ and $\setMO$ relations.
	
	\item \cds does not halt at the detection of the first buggy 
	trace; instead, it continues to provide all buggy traces as 
	required by \ourtechnique.
%
\end{enumerate}
To bridge the gap between \cds's output and our requirements,
we modify \cds's code to accept program location as an attribute of the 
program events and to halt at the first buggy trace when specified.
\ourtechnique and \fasttechnique are available as an
open-source tool that performs fence synthesis for \cc programs 
at:~\url{https://github.com/singhsanjana/fensying}.

\noindent
{\bf Experimental setup.}
The experiments were performed on an Intel(R) Xeon(R) CPU E5-1650 v4 @ 
3.60GHz with 32GB RAM and 32 cores.
We collected a set of 1389  litmus tests of buggy \cc input programs 
(borrowed from Tracer~\cite{tracer2018})
to validate the correctness of \ourtechnique and
\fasttechnique experimentally.
We study the performance of \ourtechnique and \fasttechnique on a set of 
benchmarks borrowed from previous works on model checking under the \cc 
memory model and its 
variants~\cite{abdulla2019verification,tracer2018,cds,singh2021dynamic}.

\noindent
{\bf Experimental validation.}
\begin{table}[t]
	\caption{Litmus Testing Summary}
	\centering
	
	{\em Litmus Tests Summary}
	\setlength{\tabcolsep}{2pt}
	\begin{tabular}{|l|l|l|l|l|l|l|l|l|l|}
		\hline
		Tests & min-BT & max-BT & avg-BT & min-syn & max-syn & avg-syn & min-str & max-str & avg-str \\
		\hline
		1389  & 1      & 9      & 1.05   & 1       & 4       & 2.25    & 0       & 0       & 0       \\		
		\hline
		
		\multicolumn{10}{r}{BT: \#buggy traces, syn: \#fences synthesized, str: \# fences strengthened} \\
		\multicolumn{10}{r}{min: minimum, max: maximum, avg: average} \\
	\end{tabular}

	{\em Results Summary}
	\setlength{\tabcolsep}{3pt}
	\begin{tabular}{|l|r|r|r|r|r|r|}
		\hline
		               & completed (syn+no fix) & TO & NO & Tbtg (total) & TF (total) & Ttotal   \\
		\hline
		\ourtechnique  & 1333 (1185+148)        & 56 & 0  & 50453.19     &	36896.06  & 87266.09 \\ 
		\fasttechnique & 1355 (1207+148)        & 34 & 0  & 30703.71     & 49068.61   & 79772.32 \\ 
		\hline
		
		\multicolumn{2}{l}{Times in seconds.} & \multicolumn{5}{r}{TO: 15min for {\texttt \scriptsize BTG} + 15min for technique} \\
		
		\multicolumn{7}{l}{T{\texttt \scriptsize btg}: Time of {\texttt \scriptsize BTG}, TF: Time of \ourtechnique or \fasttechnique, Ttotal: T{\texttt \scriptsize btg}+TF}
	\end{tabular}
	\label{tab:litmus}
\end{table}

The summary of the 1389 litmus tests is shown under {\em Litmus Tests Summary}, 
Table~\ref{tab:litmus}.
%
The 
number of buggy traces for the litmus tests ranged
between 1-9 with an average of 1.05, while the number of fences synthesized
ranged between 2-4. None of the litmus tests contained fences in the input
program. Hence, no fences were strengthened in any of the tests.

We present the results of \ourtechnique and \fasttechnique under {\em Result
Summary}, Table~\ref{tab:litmus}. The results have been averaged over five runs for
each test. 
\fasttechnique 
timed out  (column `TO') on a fewer number of tests (34 tests) in comparison to 
\ourtechnique (56 tests).
The techniques could not fix 148 tests with \cc fences (`no fix').
%
%
The column `NO' represents the number of tests where the fences synthesized or 
strengthened is nonoptimal. 
To report the values of `NO', we conducted a sanity test on the fixed program 
as follows:
we create versions $P_1, ... P_k$ of the fixed program $\fx{P}$ \st in each 
version, one of the fences of $\fx{P}$ is either weakened or eliminated. Each
version is then tested separately on \btg. The sanity check is successful 
if a buggy trace is returned for each version.

\noindent
{\bf Performance analysis.}
\noindent
\begin{table}[t]
	\caption{Comparative performance analysis}
	\centering
	\scriptsize
	\setlength{\tabcolsep}{2pt}
	\begin{tabular}{|l|l|r|lrrr|llrrr|}
		\hline
		& & & \multicolumn{4}{|c|}{\ourtechnique} & \multicolumn{5}{|c|}{\fasttechnique} \\
		Id & Name & \#BT & syn+str & Tbtg & TF & Ttotal & iter & syn+str & Tbtg & TF & Ttotal \\
		\hline
		1 & peterson(2,2) & 30 & 1+0 & 2.63 & 54.31 & 56.94 & 1:1 & 1:1+0:0 & 0.18 & 2.07 & 2.25 \\
		2 & peterson(2,3) & 198 & 1+0 & 29.96 & 594.34 & 624.3 & 1:1 & 1:1+0:0 & 0.53 & 3.58 & 4.11 \\
		3 & peterson(4,5) & ? & $-$ & $-$ & \fto & $-$ & 1:1 & 1:1+0:0 & 397.51 & 21.07 & 418.58 \\
		4 & peterson(5,5) & ? & $-$ & \bto & $-$ & $-$ & 1:1 & 1:1+0:0 & \bto & 31.52 & *931.52 \\
		\hline
		
		5 & barrier(5) & 136 & 1+0 & 1.09 & 207.74 & 208.83 & 1:1 & 1:1+0:0 & 0.13 & 1.40 & 1.53 \\
		6 & barrier(10) & 416 & 1+0 & 3.37 & 565.44 & 568.81 & 1:1 & 1:1+0:0 & 0.2 & 2.70 & 2.9 \\
		7 & barrier(100) & 31106 & $-$ & $-$ & \fto & $-$ & 1:1 & 1:1+0:0 & 34.2 & 198.54 & 232.74 \\
		8 & barrier(150) & ? & $-$ & $-$ & \fto & $-$ & 1:1 & 1:1+0:0 & 117.09 & 399.20 & 516.29 \\
		9 & barrier(200) & $-$ & $-$ & $-$ & $-$ & $-$ & $-$ & $-$ & $-$ & \fto & $-$ \\
		\hline
		
		10 & store-buffer(2) & 6 & 2+0 & 0.08 & 0.91 & 0.99 & 1:1 & 2:2+0:0 & 0.04 & 0.05 & 0.09 \\
		11 & store-buffer(4) & 20 & 2+0 & 1.61 & 195.35 & 196.96 & 1:1 & 2:2+0:0 & 1.20 & 0.05 & 1.25 \\
		12 & store-buffer(5) & 30 & $-$ & $-$ & \fto & $-$ & 1:1 & 2:2+0:0 & 14.07 & 0.22 & 14.29 \\
		13 & store-buffer(6) & 42 & $-$ & $-$ & \fto & $-$ & 1:1 & 2:2+0:0 & 171.09 & 0.15 & 171.24 \\
		14 & store-buffer(10) & ? & $-$ & \bto & $-$ & $-$ & 1:1 & 2:2+0:0 & \bto & 0.05 & *900.05 \\
		\hline
		
		15 & dekker(2) & 54 & 2+0 & 0.17 & 0.27 & 0.44 & 1:1 & 2:2+0:0 & 0.26 & 0.04 & 0.3 \\
		16 & dekker(3) & 1596 & $-$ & $-$ & \fto & $-$ & 1:1 & 2:2+0:0 & 586.46 & 1.34 & 587.8 \\
		\hline
		
		17 & dekker-fen(2,3) & 54 & 1+1 & 0.15 & 0.29 & 0.44 & 1:1 & 1:1+1:1 & 0.25 & 0.05 & 0.3 \\
		18 & dekker-fen(3,2) & 730 & $-$ & $-$ & \fto & $-$ & 1:1 & 1:1+1:1 & 159.84 & 5.56 & 165.4 \\
		19 & dekker-fen(3,4) & 3076 & $-$ & \bto & $-$ & $-$ & 1:1 & 1:1+1:1 & \bto & 6.06 & *906.06 \\
		\hline
		
		20 & burns(1) & 36 & $-$ & $-$ & \fto & $-$ & 7:8 & 8:10+2:2 & 0.61 & 4.69 & 5.3 \\
		21 & burns(2) & 10150 & $-$ & $-$ & \fto & $-$ & 6:7 & 8:10+0:1 & 71.53 & 554.6 & 626.13 \\
		22 & burns(3) & ? & $-$ & \bto & $-$ & $-$ & $-$ & $-$ & $-$ & \fto & $-$ \\
		\hline
		
		23 & burns-fen(2) & 100708 & $-$ & $-$ & \fto & $-$ & 5:7 & 4:6+3:3 & 329.41 & 43.96 & 373.37 \\
		24 & burns-fen(3) & ? & $-$ & \bto & $-$ & $-$ & 5:7 & 4:6+3:3 & \bto & 70.14 & *970.14 \\
		\hline
		
		25 & linuxrwlocks(2,1) & 10 & $-$ & $-$ & \fto & $-$ & 1:1 & 2:2+0:0 & 0.13 & 0.12 & 0.25 \\
		26 & linuxrwlocks(3,8) & 353 & $-$ & $-$ & \fto & $-$ & 2:2 & 3:4+0:0 & 686.52 & 0.41 & *686.93 \\
		\hline
		
		27 & seqlock(2,1,2) & 500 & $-$ & $-$ & \fto & $-$ & 1:1 & 1:1+0:0 & 341.54 & 2.38 & 343.92 \\
		28 & seqlock(1,2,2) & 592 & $-$ & $-$ & \fto & $-$ & 1:2 & 1:2+0:0 & 119.88 & 27.69 & 147.57 \\
		29 & seqlock(2,2,3) & ? & $-$ & \bto & $-$ & $-$ & 1:2 & 1:2+0:0 & \bto & 88.52 & 988.52* \\
		\hline
		
		30 & bakery(2,1) & 6 & 1+0 & 0.25 & 25.42 & 2.88 & 1:1 & 1:1+0:0 & 0.07 & 0.18 & 0.25 \\
		31 & bakery(4,3) & 7272 & $-$ & $-$ & \fto & $-$ & 1:1 & 1:1+0:0 & 166.11 & 5.68 & 171.79 \\
		32 & bakery(4,4) & 50402 & $-$ & $-$ & \fto & $-$ & 1:1 & 1:1+0:0 & \bto & 18.17 & 918.17* \\
		\hline
		
		33 & lamport(1,1,2) & 1 & No fix. & 0.06 & 0.05 & 0.11 & 1:1 & No fix. & 0.04 & 0.05 & 0.09 \\
		34 & lamport(2,2,1) & 1 & No fix. & 411.94 & 0.05 & 411.99 & 1:1 & No fix. & 53.34 & 0.05 & 53.39 \\
		35 & lamport(2,2,3) & ? & $-$ & \bto & $-$ & $-$ & 1:1 & No fix. & 389.77 & 0.05 & 389.82 \\
		\hline
		
		36 & flipper(5) & 297 & 2+0 & 6.22 & 254.18 & 260.40 & 1:1 & 2+0 & 2.51	& 0.02 & 2.53 \\
		37 & flipper(7) & 4493 & $-$ & $-$ & \fto & $-$ & 1:1 & 2+0 & 119.21 & 0.02 & 119.23 \\
		38 & flipper(10) & ? & $-$ & $-$ & \fto & $-$ & 1:1 & 2+0 & \bto & 0.03 & 900.03* \\
		\hline
		
		\multicolumn{12}{l}{T{\texttt \scriptsize btg}: Time of {\texttt \scriptsize BTG}, TF: Time of technique (\ourtechnique or \fasttechnique), Ttotal: T{\texttt \scriptsize btg}+TF}
	\end{tabular}
	\label{tab:bench}
\end{table}

We contrast the performance of the techniques using 
a set of benchmarks that produce buggy traces under \cc. 
The results are averaged over five runs.
 Table~\ref{tab:bench} reports the results where `\#BT' shows the
number of buggy traces, `iter' shows the minimum:maximum number of 
iterations performed by \fasttechnique over the five runs and, 
`\fto' and `\bto' represent \ourtechnique/\fasttechnique time-out and
\btg time-out, respectively (set to 15 minutes each).
A `?' in `\#BT' signifies that \btg could not scale for the test, so
the number of buggy traces is unknown.
The column (`syn+str') under \fasttechnique reports the minimum:maximum 
number of fences synthesized and/or strengthened.
We add a `*' against the time when \btg timed out in detecting that 
the fixed program has no more buggy traces.

\begin{figure}[t]
	{\centering
	\fbox{\includegraphics[scale=0.45]{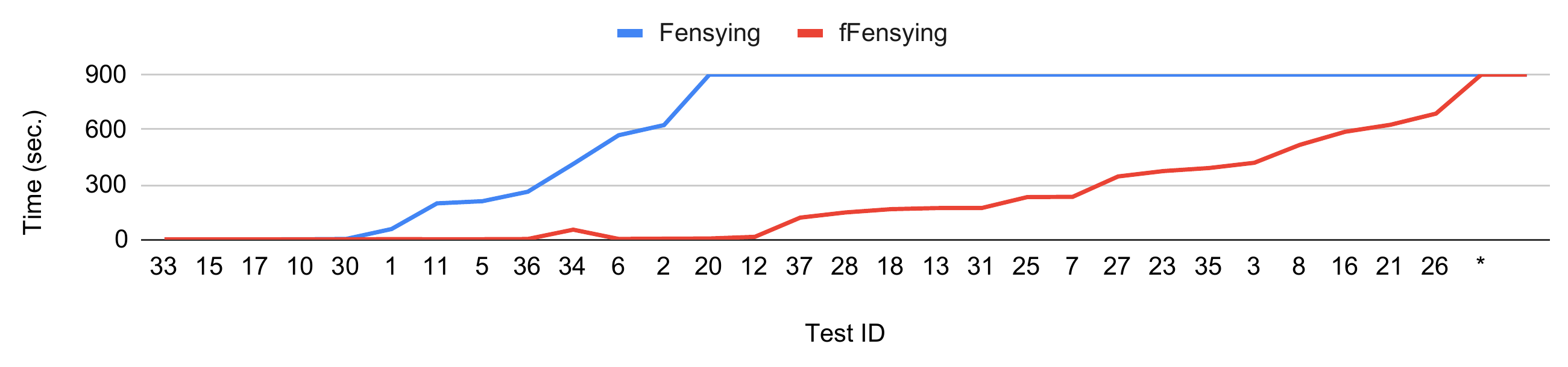}}}
	{\scriptsize $\star$ represents the remaining Test IDs (tests that timeout for both \ourtechnique and \fasttechnique)}
	\caption{Performance comparison between \ourtechnique and \fasttechnique}
	\label{fig:graph}
\end{figure}

%
The performance of \ourtechnique and \fasttechnique is diagrammatically
contrasted in Figure~\ref{fig:graph}.
It is notable that \fasttechnique significantly 
outperforms \ourtechnique in terms of the time of execution
and scalability and
adds extra fences in only 7 tests with an average of 1.57 additional fences.
With the increase in the number of buggy traces, an exponential
rise in  \ourtechnique's time leading to \fto was observed;
except in cases 12, 13, 20, and 25, where \ourtechnique times out
with as low as 10 traces.
%
%
The tests time-out in {\em Johnson's} cycle detection due to a high
density of the number of related events or the number of cycles. 
%
%

%
\fasttechnique analyzes a remarkably smaller number 
of buggy traces (`iter') in comparison with `\#BT' (${\le}2$ traces for 
$\sim$85\% of tests). We conclude that a solution corresponding to a 
single buggy trace fixes more than one buggy traces.
As a result, \fasttechnique can scale to tests with thousands of buggy traces 
and we witness an average speedup of over 67x, with over 100x speedup in 
$\sim$41\% of tests, against \ourtechnique.  
%
%
%

\noindent{\em Interesting cases.}
Consider test 16, where \btg times out in 3/5 runs and completes in
$\sim$100s in the remaining 2 runs. A fence is synthesized between two
events, $e_1$ and $e_2$, that are inside a loop. Additionally, $e_1$ is
within a condition. Depending on where the fence is synthesized (within
the condition or outside it), \btg either runs out of time or finishes
quickly.
Similarly, \btg for test 26 times out in 3/5 runs. However, the reason here
is the additional nonoptimal fences synthesized that increase the analysis
overhead of the chosen \btg (\cds).

Note that, for most benchmarks, \fasttechnique's scalability is 
limited by \bto and observably \fasttechnique's time is much 
lesser than \fto for such cases.
Therefore, an alternative \btg would significantly improve
\fasttechnique's performance.

%

\section{Related Work} \label{sec:related}
The literature on fence synthesis is rich with 
techniques targeting the x86-TSO
\cite{abdulla2012automatic,abdulla2015best,alglave2014don,alglave2010fences,bender2015declarative} 
and sparc-PSO 
\cite{abdulla2015precise,linden2013verification} 
memory models or both
\cite{joshi2015property,kuperstein2012automatic,meshman2014synthesis}.
%
The work in~\cite{bender2015declarative} and
\cite{kuperstein2012automatic} perform fence synthesis for ARMv7 and 
RMO memory models.
The works in~\cite{abdulla2015precise,alglave2010fences,fang2003automatic}
are proposed for Power memory model, where~\cite{fang2003automatic} also 
supports IA-32 memory model.

Most fence synthesis techniques introduce additional 
ordering in the program events with the help of fences
\cite{abdulla2015best,alglave2014don,alglave2010fences,bender2015declarative,fang2003automatic,joshi2015property,linden2013verification,meshman2014synthesis,taheri2019polynomial}.
However, the axiomatic definition of ordering varies with 
memory models.
As a consequence, most existing techniques (such as those for TSO 
and PSO) may not detect \cc buggy traces due to a strong implicit 
ordering.
While the techniques~\cite{alglave2014don,alglave2010fences,taheri2019polynomial}
are parametric in or oblivious to the memory model,
they introduce ordering between {\em pairs} of events that is
{\em globally visible} (to all threads).
Such an ordering constraint is restrictive for weaker models
such as \cc that may require ordering on a {\em set} of events
that may be {\em conditionally visible} to a thread.
Similarly, \cite{joshi2015property} proposes a bounded technique
applicable to any memory model that supports interleaving with
reordering. Program outcomes under \cc may not be
feasible under such a model.
Moreover, any existing technique, cannot fix a \cc input program 
while conserving its portability.

%
%
%
%
%
Some earlier works such as 
\cite{abdulla2012automatic,alglave2010fences,fang2003automatic} 
synthesize fences to restrict outcomes to SC or its variant for store-buffering 
\cite{abdulla2015best}. 
%
Most fence synthesis techniques 
\cite{abdulla2015precise,joshi2015property,kuperstein2012automatic,linden2013verification,meshman2014synthesis}
attempt to remove traces violating a safety property specification
under their respective axiomatic definition of memory model. 
Various works  \cite{abdulla2015precise,abdulla2015best,bender2015declarative,joshi2015property,kuperstein2012automatic,meshman2014synthesis,taheri2019polynomial}
perform optimal fence synthesis where the optimality (in the absence of types of fences) is simply defined as the smallest set of fences.
Technique~\cite{alglave2014don} assigns weights to various types 
of fences (similar to our work) and defines optimality on the summation 
of fence weights of candidate solutions.
However, their definition of optimality is incomparable with ours, and
no prior work establishes the advantage of one definition over the other.

%
Lastly, a recent technique~\cite{oberhauser2021vsync} fixes a buggy \cc
program by strengthening memory access events instead of synthesizing fences.

\section{Conclusion and Future Work} \label{sec:conclusion}
This paper proposed the first fence synthesis techniques for \cc programs:
an optimal  (\ourtechnique) and a near-optimal
 (\fasttechnique).
The work also presented theoretical arguments that showed the
correctness of the synthesis techniques. The experimental validation
demonstrated the effectiveness of \fasttechnique vis-\`{a}-vis
optimal \ourtechnique.
%
%
As part of future work, we will investigate extending the
presented methods (i) to support richer constructs such as locks
and (ii) to include strengthening memory accesses to fix buggy traces.

%
%
\bibliographystyle{splncs03}
\bibliography{References}

\end{document}